\documentclass[fleqn,usenatbib]{mnras}

\usepackage{newtxtext,newtxmath}
\usepackage[T1]{fontenc}
\usepackage{booktabs}
\usepackage{tabularx}
\usepackage{subcaption}
\usepackage{float}
\usepackage{longtable}
\usepackage{array}
\usepackage{graphicx}
\usepackage{amsmath}
\usepackage{multirow}
\usepackage{comment}
\usepackage{placeins}

\usepackage{graphicx}	
\usepackage{amsmath}	
\usepackage{scalerel,tikz}
\usetikzlibrary{svg.path,arrows.meta,calc}
\definecolor{orcidlogocol}{HTML}{A6CE39}
\tikzset{orcidlogo/.pic={
		\fill[orcidlogocol] svg{M256,128c0,70.7-57.3,128-128,128C57.3,256,0,198.7,0,128C0,57.3,57.3,0,128,0C198.7,0,256,57.3,256,128z};
		\fill[white] svg{M86.3,186.2H70.9V79.1h15.4v48.4V186.2z}
		svg{M108.9,79.1h41.6c39.6,0,57,28.3,57,53.6c0,27.5-21.5,53.6-56.8,53.6h-41.8V79.1z M124.3,172.4h24.5c34.9,0,42.9-26.5,42.9-39.7c0-21.5-13.7-39.7-43.7-39.7h-23.7V172.4z}
		svg{M88.7,56.8c0,5.5-4.5,10.1-10.1,10.1c-5.6,0-10.1-4.6-10.1-10.1c0-5.6,4.5-10.1,10.1-10.1C84.2,46.7,88.7,51.3,88.7,56.8z};
}}
\newcommand\orcidicon[1]{\href{https://orcid.org/#1}{\mbox{\scalerel*{
				\begin{tikzpicture}[yscale=-1,transform shape]
					\pic{orcidlogo};
				\end{tikzpicture}
			}{|}}}}

\usepackage{hyperref}

\title[Limited impact of source stochasticity]{Galaxy-to-galaxy stochasticity has a limited impact on reionization and the 21-cm signal at fixed mean emissivity}

\author[Zhu \& Qin]{
Shihai Zhu~\orcidicon{0009-0001-6409-0012}$^{1}$ and Yuxiang Qin~\orcidicon{0000-0002-4314-1810}$^{1}$\thanks{E-mail: Yuxiang.L.Qin@gmail.com}\\
$^{1}$Research School of Astronomy and Astrophysics, Australian National University, Canberra, ACT 2611, Australia\\
}

\date{}
\pubyear{\the\year{}}

\begin{document}
\label{firstpage}
\pagerange{\pageref{firstpage}--\pageref{lastpage}}
\maketitle

\begin{abstract}
Galaxy-to-galaxy stochasticity may be less important for reionization, IGM
heating, and the 21-cm signal than commonly assumed. 
We find that most of its apparent impact arises because
introducing or removing scatter changes the mean emissivity of the galaxy
population, rather than from redistributing radiation at fixed mean
emissivity. We use the
\textsc{Meraxes} semi-analytic model to separate these effects. For
ionising sources, we vary the scatter in the escape
fraction and in the SFR--halo mass relation across three source
prescriptions. For the X-ray heating sources, we introduce luminosity scatter and also
separately remove SFR scatter. Without
recalibration, these modifications shift the mean emissivities, changing the
timing of reionization and IGM heating and producing substantial differences
in the 21-cm signal. Once the relevant mean emissivities are restored,
however, the global ionizing and thermal histories converge and the power-spectrum differences
fall below the forecast sensitivity of a deep {\it SKA}-Low observation.
There are some weak changes in ionisation morphology, depending on which galaxies dominate the ionising emissivity and on their
response to photoheating. Within the
models and scales considered here, calibrating the mean radiative
emissivities is therefore substantially more important than resolving their
galaxy-to-galaxy scatter.
\end{abstract}

\section{Introduction}

Cosmic dawn and the epoch of reionization (EoR) mark the transformation of
the intergalactic medium (IGM) by the first generations of stars and galaxies.
Ly\(\alpha\) photons couple the hydrogen spin temperature to the gas kinetic
temperature, X-rays heat and weakly ionise the predominantly neutral IGM, and
Lyman-continuum photons ultimately reionize it
\citep{Zaroubi2013EoR,PritchardFurlanetto2007,
Mesinger2011_21cmFAST,Pacucci2014XraySpectra}. Together, these processes set
the contrast of the redshifted 21-cm signal against the cosmic microwave
background (CMB). Its sky-averaged evolution traces the progression from
coupling through heating to reionization, while its spatial fluctuations
encode the distributions of the sources and the resulting Ly\(\alpha\),
temperature, and ionisation fields.

Because galaxies are biased tracers of the density field, these radiation
fields are spatially inhomogeneous. Ionised regions grow around clustered
sources before overlapping, whereas inhomogeneous X-ray heating generates
temperature fluctuations in the surrounding neutral gas
\citep{Furlanetto2004GrowthHII,PritchardFurlanetto2007,
McQuinn2007Morphology,Becker2015PatchyReionization}. The CMB optical depth
constrains the integrated ionisation history
\citep{PlanckCollaboration2020CosmoParams}, while quasar and galaxy
observations indicate that reionization was largely complete by
\(z\sim5.4\), remained highly inhomogeneous near its end, and that the IGM
was still significantly neutral at \(z\sim7\)
\citep{Becker2015PatchyReionization,McGreer2015endreionization,
Greig2017J1120,Mason2018ReionizingZ7,Bosman2022,Qin2025,Zhu2022LyB}.
These measurements constrain the ionisation history but do not uniquely
determine the galaxies responsible for it or the preceding thermal history.
The 21-cm signal provides a complementary probe of both.

Numerical models are required to connect the galaxy population to the
coupled ionisation, thermal, and 21-cm fields. Radiation-hydrodynamic
simulations can self-consistently follow galaxy formation, radiative
transport, and the response of the IGM. They reveal substantial spatial and
temporal variations in source properties such as the escape fraction
\citep{Paardekooper2015FiBY,KimmCen2014Fesc,Trebitsch2017Fesc,
Ma2015FIREFesc,Yeh2023ThesanFesc}, but their computational cost limits the
volumes, resolution, and parameter ranges that can be explored.
Semi-analytic and semi-numerical methods instead apply approximate source
prescriptions to dark-matter halo fields, enabling larger volumes and
systematic parameter studies
\citep{Park2019,Mutch2016DRAGONSIII,Hutter2021AstraeusI}.
These prescriptions often compress the galaxy population into deterministic
or mean galaxy--halo relations, neglecting part of the galaxy-to-galaxy
stochasticity present in more detailed calculations. Whether this omission
materially affects reionization and 21-cm predictions after the mean source
relations have been calibrated remains unclear.

Schematically, the escaped ionising emissivity can be written as
\begin{equation}
    \dot n_{\rm ion}
    \propto
    \int {\rm d}M_{\rm h}\,
    \frac{{\rm d}n_{\rm h}}{{\rm d}M_{\rm h}}\,
    f_{\rm duty}(M_{\rm h},z)\,
    \left\langle {\rm SFR}\,N_\gamma\,f_{\rm esc} \right\rangle,
    \label{eq:intro_ionising_budget}
\end{equation}
where \({\rm d}n_{\rm h}/{\rm d}M_{\rm h}\) is the halo mass
function, \(f_{\rm duty}(M_{\rm h},z)\) is the fraction of haloes
hosting active star formation, SFR is the star-formation rate,
\(N_\gamma\) is the number of ionising photons produced per
stellar baryon, and \(f_{\rm esc}\) is the fraction of ionising
photons that escape into the IGM. The angle brackets denote an
average over active sources at fixed halo mass and redshift.
The cumulative escaped photon production is obtained by integrating this
emissivity over time. Stochasticity has two conceptually distinct effects.
First, scatter can change the arithmetic mean emissivity even when it is
introduced about an unchanged median relation, particularly for non-linear
or log-normal source prescriptions \citep{Nikolic2024Stochasticity}.
Second, at fixed mean emissivity, it redistributes radiation among galaxies,
halo masses, and environments. For ionising photons, this redistribution can
change the clustering of the effective sources and the morphology of
reionization \citep{Furlanetto2006CharacteristicScales,
McQuinn2007Morphology,Hassan2022SIMBA}; for X-rays, it can alter the spatial
pattern of IGM heating and hence the spin-temperature fluctuations
\citep{PritchardFurlanetto2007,Davies2025DiscreteGalaxies}. The relative
importance of these effects is not obvious. The radiation fields are
generated by the collective emission of many galaxies, and X-ray propagation
additionally smooths the heating field over extended distances. Substantial
galaxy-to-galaxy stochasticity therefore need not produce a large change in
the ionisation or thermal fields once the corresponding mean radiative
emissivity is fixed.

The source properties determining these emissivities are all expected to vary among
galaxies. Differences in halo assembly, gas accretion, star formation, and
feedback generate scatter in stellar properties at fixed halo mass
\citep{Moster2013SHMR,Hopkins2014FIRE,Allen2019SMHMscatter,
Pei2024SHMRScatter}. The escape fraction depends on the small-scale coupling
between young stars, stellar feedback, and the multiphase interstellar
medium, and can vary strongly between galaxies and over time
\citep{Paardekooper2015FiBY,Trebitsch2017Fesc,Ma2015FIREFesc,
KimmCen2014Fesc,Yeh2023ThesanFesc}. The X-ray luminosity at fixed SFR also
varies because metallicity affects the formation and evolution of high-mass
X-ray binaries, whose finite populations introduce an additional stochastic
component \citep{Mineo2012HMXB,Lehmer2021HMXBLF}. Moreover, source
stochasticity can propagate beyond an instantaneous change in the radiation
field. In a coupled galaxy--reionization calculation, modifications to the
ionisation and ultraviolet-background (UVB) fields regulate later gas
accretion and star formation in low-mass haloes
\citep{Efstathiou1992PhotoionizationDwarfs,
SobacchiMesinger2013UVBfeedback}.

Using a semi-empirical model that sampled stochasticity in several galaxy
properties, \citet{Nikolic2024Stochasticity} found that neglecting
galaxy-to-galaxy variations can bias the mean and variance of the ionising
emissivity and delay the inferred completion of reionization by
\(\Delta z\sim1{-}2\). This demonstrates that adding scatter about an
unchanged median relation can be important because it shifts the mean
emissivity. It does not, however, establish that stochasticity itself leaves
a comparable imprint once this mean is restored. The calculation also did
not evolve galaxy formation together with the spatially inhomogeneous
ionisation and UVB fields. It therefore remains unclear whether
redistributing a fixed escaped emissivity among galaxies appreciably changes
reionization morphology. An analogous question applies to cosmic heating:
at fixed mean X-ray emissivity, does redistributing luminosity among discrete
galaxies measurably alter the global or fluctuating 21-cm signal?

In this work, we address these questions using the semi-analytic model
\textsc{Meraxes}\footnote{\url{https://github.com/meraxes-devs}}, which couples the evolving galaxy population to spatially
inhomogeneous reionization and photoheating feedback
\citep{Mutch2016DRAGONSIII}. For the UV-ionising source population, we
introduce scatter in \(f_{\rm esc}\) and separately remove the intrinsic
scatter in the SFR--halo mass relation, considering three escape-fraction
prescriptions. We also perform a separate set of calculations with the full
spin-temperature field, either introducing \(0.5\)-dex scatter in X-ray
luminosity at fixed SFR or removing the intrinsic SFR
scatter at fixed halo mass. In each case, we
compare uncalibrated models, which
retain the scatter-induced change in the mean radiative emissivities, with
recalibrated models that restore the corresponding baseline means. This
design separates the mean-emissivity effect from the residual effect of
redistributing radiation among sources and environments.

We find that galaxy-to-galaxy stochasticity is not a major driver in either
application once the relevant mean emissivity is calibrated. Most of the
apparent response instead follows from changes in the mean radiative
emissivity. Once the escaped ionising
emissivity is controlled, the global reionization histories nearly converge
and the residual changes in
ionisation morphology and the 21-cm power spectrum are generally small. A
stronger morphological response appears only when low-mass galaxies dominate
the ionising emissivity, for which source redistribution couples to subsequent
UVB feedback. Likewise, restoring the mean X-ray emissivity and, for the
no-SFR-scatter model, the mean stellar Ly\(\alpha\) emissivity removes the
shifts in coupling and heating produced by the source modifications, leaving small residual
differences in the global 21-cm signal and power spectrum.

We organise the paper as follows. In Section~\ref{sec:method}, we describe
the \textsc{Meraxes} model, the UV and X-ray source prescriptions, the model
suite, and the reionization and 21-cm diagnostics. In
Section~\ref{sec:results}, we separate the mean-emissivity and redistribution
effects on the global histories, ionisation morphology, and 21-cm signal,
including their coupling to UVB feedback. We summarise our conclusions in
Section~\ref{sec:conclusion}. Throughout this work, we adopt a flat
\(\Lambda\)CDM cosmology with \(h=0.6751\), \(\Omega_{\rm m}=0.3121\),
\(\Omega_{\rm b}=0.0490\), \(\Omega_\Lambda=0.6879\),
\(\sigma_8=0.815\), and \(n_{\rm s}=0.968\).

\section{Model and methodology}
\label{sec:method}

We first describe the simulation and the coupled galaxy-formation and
reionization model. We then introduce the stochastic UV and X-ray source
prescriptions, define the uncalibrated and recalibrated model variants, and
present the diagnostics used to quantify their effects on reionization,
IGM heating, and the 21-cm signal.

\subsection{Simulation and the \textsc{Meraxes} model}
\label{sec:meraxes}

We use halo merger trees generated from the
\texttt{Genesis\_L210\_N4320} dark-matter simulation, which follows
\(4320^3\) collisionless particles in a periodic box of side length
\(210\,h^{-1}\,{\rm cMpc}\), with a particle mass of
\(9.95\times10^{6}\,h^{-1}\,{\rm M_\odot}\). Haloes are identified using
\textsc{VELOCIraptor} \citep{Elahi2019VELOCIraptor}. The resulting halo
catalogues and merger trees are augmented following
\citet{Balu2023ThermalReionisationHistory} to resolve the atomic-cooling
halo population at \(z\leq20\). By linking each halo to its progenitors and
descendants across successive snapshots, the merger trees encode its
hierarchical assembly history and provide the backbone for modelling the
evolution of its galaxies.

Galaxy formation is modelled using the semi-analytic model
\textsc{Meraxes} \citep{Mutch2016DRAGONSIII}. Unlike conventional
semi-analytic models that evolve each merger tree independently,
\textsc{Meraxes} advances the full galaxy population simultaneously from
one snapshot to the next. This horizontal traversal allows the ionisation
and ultraviolet-background (UVB) fields generated at one snapshot to
regulate subsequent galaxy evolution following
\citet{SobacchiMesinger2013UVBfeedback}.

Within each halo, newly accreted gas enters a hot reservoir and can
subsequently cool onto its galaxies, supplying the cold gas available for
star formation. Stellar evolution returns mass and metals to the gas
reservoirs, while supernova feedback reheats or ejects gas and regulates the
build-up of stellar mass and star formation, particularly in low-mass
haloes. Galaxy mergers redistribute gas and stars and can trigger additional
star formation and rapid growth of the central supermassive blackhole
\citep{Qin2017MNRAS.472.2009Q}. These processes determine the stellar masses
and SFRs used to construct the radiative source populations. Their
prescriptions and free parameters follow \citet{Qin2023MNRAS.526.1324Q} and
are held fixed throughout this work. The resulting galaxy population
reproduces the observed rest-frame UV luminosity functions over
\(5\lesssim z\lesssim13\), including the pre-{\it JWST} constraints at
\(z\lesssim10\) and the faint end of early {\it JWST} measurements at higher
redshift \citep[e.g.][]{Bouwens2021AJ....162...47B,Harikane2023ApJS..265....5H}.

At each snapshot, the galaxy source properties and positions are mapped onto
a \(512^3\) Cartesian grid, corresponding to a cell width of
\(0.410\,h^{-1}\,{\rm cMpc}\). The UV-ionising source field is used to
calculate the three-dimensional neutral-fraction field,
\(x_{\rm HI}(\boldsymbol{x})\), and the local UVB. Once a region is ionised,
UVB photoheating suppresses baryonic infall into low-mass haloes, reducing
their later gas supply and star formation. Throughout this work,
``feedback'' refers specifically to this UVB photo-suppression channel; all
other galaxy-formation prescriptions in \textsc{Meraxes} are held fixed.

\subsection{Escape-fraction prescriptions and stochasticity}
\label{sec:fesc}

The escape fraction, \(f_{\rm esc}\), specifies the fraction of ionising
photons that escape from galaxies into the IGM. Since \textsc{Meraxes}
does not resolve the small-scale interstellar and circumgalactic
structure that regulates photon escape, we adopt an effective
halo-mass-dependent prescription:
\begin{equation}
    f_{\rm esc}(M_{\rm vir})
    = f_{\rm esc,0}
    \left(\frac{M_{\rm vir}}{M_0}\right)^{\alpha_{\rm esc}},
    \label{eq:fesc_mass_dependence}
\end{equation}
where \(M_0=10^{10}\,{\rm M}_\odot\), and the resulting escape fraction
is restricted to \(0\leq f_{\rm esc}\leq1\). We adopt a
halo-mass-independent escape fraction as our fiducial model, with
\(\alpha_{\rm esc}=0\), and explore two alternatives that weight the
escaped ionising photon budget towards higher-mass haloes
(\(\alpha_{\rm esc}=0.5\)) or lower-mass haloes
(\(\alpha_{\rm esc}=-0.5\)). The normalisation \(f_{\rm esc,0}\) is
chosen separately for each baseline with UVB feedback to give a common
reionization midpoint. Their CMB optical depths range from \(0.051\)
to \(0.065\), consistent with
\citet{PlanckCollaboration2020CosmoParams} and the re-analysis by
\citet{Qin2020ReionizationInference} using physical reionization models.

To introduce galaxy-to-galaxy stochasticity\footnote{We perturb
\(f_{\rm esc}\) directly rather than linking it to another galaxy property
with intrinsic scatter at fixed halo mass
\citep{Seiler2019FescSKA,Qin2026arXiv260725722Q}. This empirical treatment
avoids imposing a specific and uncertain correlation and allows the
stochasticity in \(f_{\rm esc}\) to be varied independently of the adopted
galaxy--halo relation.},
we draw \(f_{\rm esc}\) from a log-normal distribution about the relation
in Equation~\eqref{eq:fesc_mass_dependence}. We adopt
\(\sigma_{\rm esc}=0.3\) dex, representative of galaxy-to-galaxy
variations suggested by potential Lyman-continuum leakers observed with
{\it JWST} \citep{Mascia2023GLASSJWST} and numerical simulations
\citep{Rosdahl2022LyCEscape,Qin2026arXiv260725722Q}. The adopted relation
defines the median of the distribution before clipping, and the scatter is
independent of halo mass. The sampled escape fractions are also restricted to
\(0\leq f_{\rm esc}\leq1\). We have also tested larger scatter, up to
\(\sigma_{\rm esc}=1.0\) dex, and find that our qualitative conclusions
are unchanged.

\begin{figure}
    \centering
    \includegraphics[width=\linewidth]{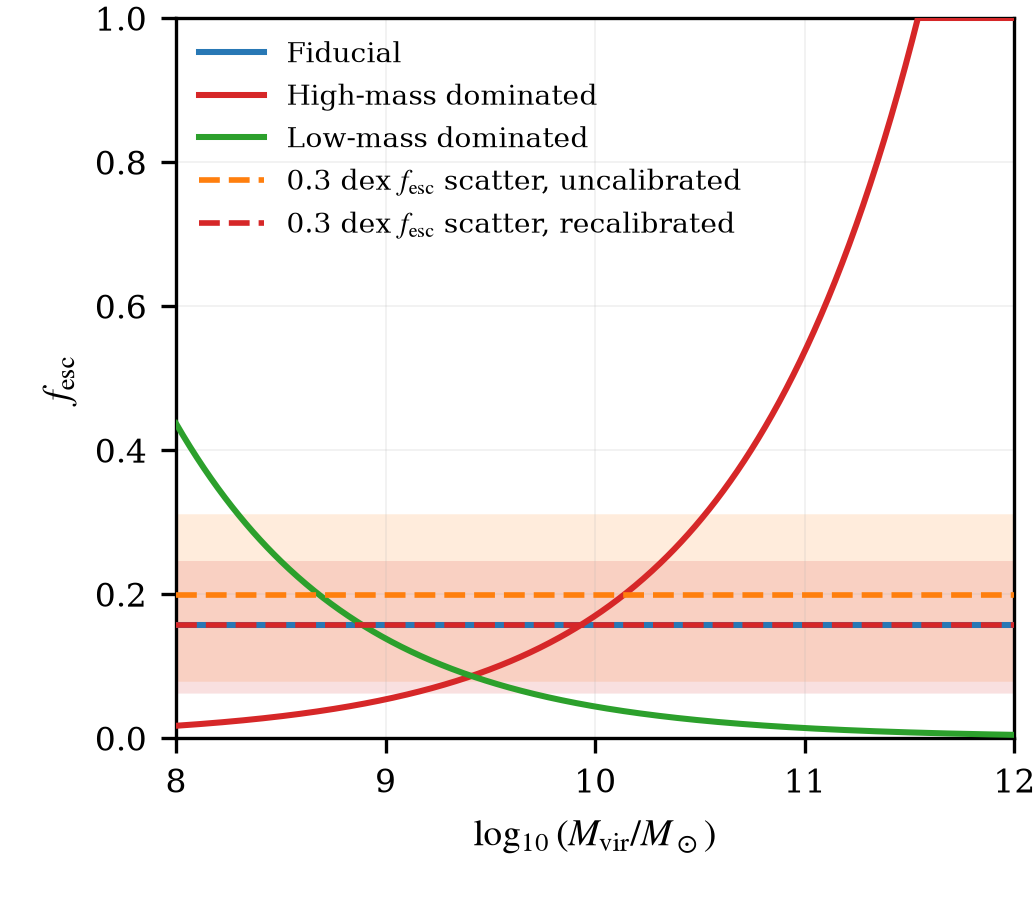}
    \caption{
    Escape fraction as a function of halo mass. Solid curves show the
    fiducial, high-mass-dominated, and low-mass-dominated baseline
    prescriptions, with \(f_{\rm esc}\) clipped at unity. For the
    fiducial prescription, dashed curves show the mean relations after
    introducing \(0.3\)-dex scatter, and shaded regions enclose the
    corresponding 16th--84th percentile ranges. The uncalibrated model
    retains the change in the escaped photon budget. The recalibrated
    model adjusts the normalisation to recover the photon budget of
    the fiducial baseline.
    }
    \label{fig:fescrelation}
\end{figure}

A distribution that is symmetric in \(\log_{10}f_{\rm esc}\) is
asymmetric in linear \(f_{\rm esc}\). Before imposing the physical bounds,
this prescription therefore increases the arithmetic mean escape
fraction\footnote{For \(\sigma_{\rm esc}=0.3\) dex, the mean is increased
by a factor of
\(\exp[\tfrac{1}{2}(\ln 10)^2\sigma_{\rm esc}^2]\simeq1.27\) before
applying the upper limit \(f_{\rm esc}\leq1\). The total ionising emissivity
need not increase by 27 per cent because \(f_{\rm esc}\) is weighted by the
halo population and its star formation, as shown in
Equation~\eqref{eq:intro_ionising_budget}. Clipping at \(f_{\rm esc}=1\)
further reduces the enhancement when the underlying escape fraction is
already high.}. We therefore consider both \emph{uncalibrated} and \emph{recalibrated}
\(f_{\rm esc}\)-scatter models, illustrated in
Fig.~\ref{fig:fescrelation}. In the uncalibrated case, scatter is applied
directly to the baseline models, retaining both the stochastic
redistribution of \(f_{\rm esc}\) among galaxies and the associated
change in the mean escaped photon budget. In the recalibrated case,
the stochastic source field is rescaled to recover the mean escaped
ionising-photon budget of the corresponding baseline model.
This removes the direct budget shift and isolates the residual effect
of redistributing ionising emission among galaxies, halo masses, and
spatial environments.

\subsection{Scatter in the SFR--halo relation}
\label{sec:shmr_scatter}

Whereas stochasticity in \(f_{\rm esc}\) is introduced explicitly, scatter
in the SFRs of galaxies arises self-consistently in \textsc{Meraxes}.
Galaxies hosted by haloes of the same mass can have different SFRs owing to
differences in their assembly and feedback histories, and therefore make
different contributions to the ionising source population. To quantify the
effect of this intrinsic variation, we compare the baseline models with
otherwise equivalent source models in which the SFR scatter at fixed halo
mass is removed.

\begin{figure}
    \centering
    \includegraphics[width=\linewidth]{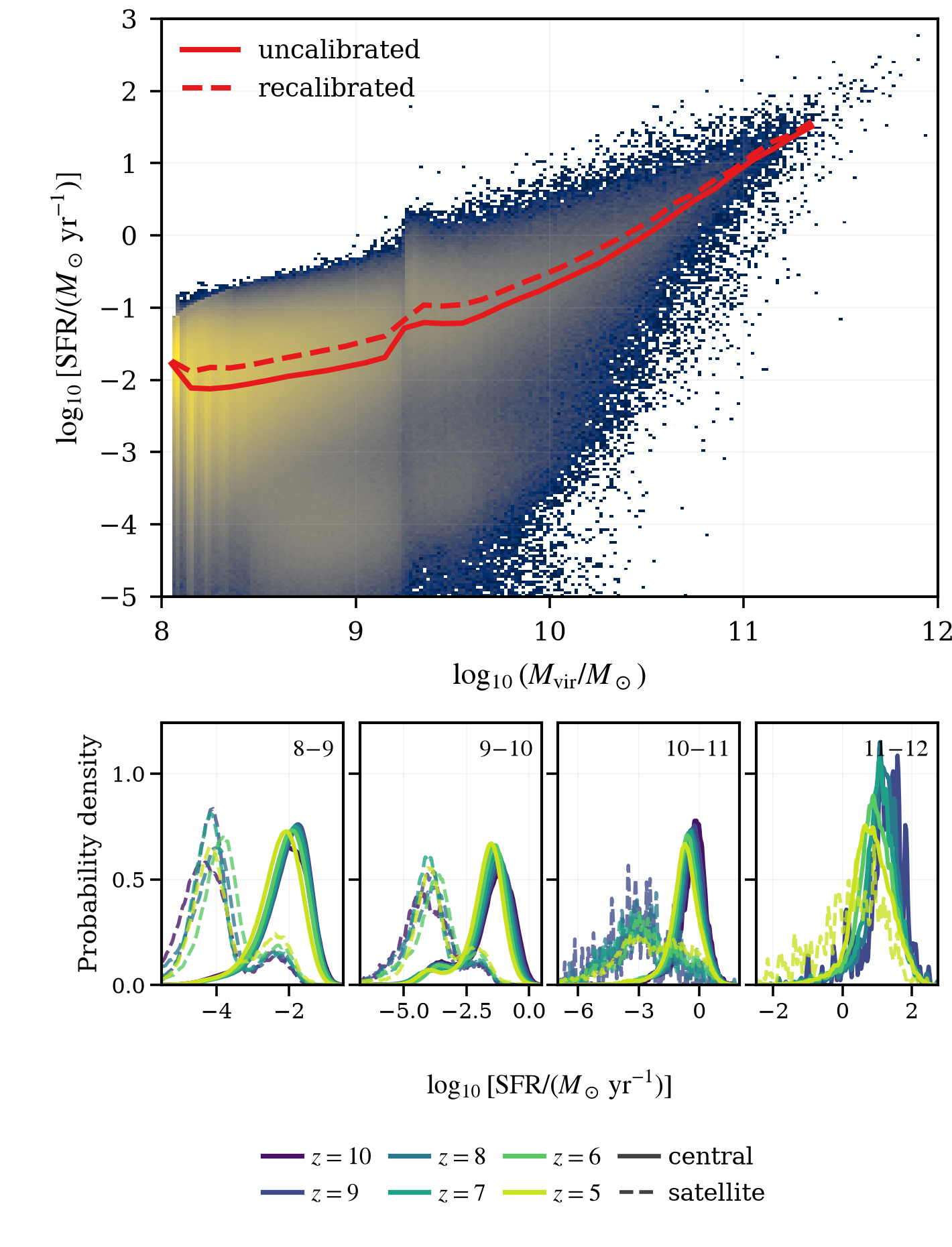}
    \caption{
    SFR distributions and deterministic source relations for the fiducial
    source family. The upper panel shows central galaxies in the baseline
    SFR--\(M_{\rm vir}\) relation at \(z\sim8\). The solid and dashed red
    curves show the uncalibrated and recalibrated no-SFR-scatter relations,
    respectively. The former is the median SFR at fixed halo mass, galaxy
    type, and redshift; recalibration rescales its source contribution to
    recover the corresponding baseline escaped photon budget. The lower
    panels show the conditional distributions of \(\log_{10}{\rm SFR}\) for
    actively star-forming galaxies in four
    \(\log_{10}(M_{\rm vir}/{\rm M}_\odot)\) intervals. Colours denote
    redshift from \(z=10\) to \(z=5\), while solid and dashed curves denote
    central and satellite galaxies, respectively.
    }
    \label{fig:source_relations}
\end{figure}

Figure~\ref{fig:source_relations} illustrates the intrinsic scatter in
the SFR--halo mass relation. The upper panel shows the distribution
of central galaxies at \(z\sim8\), and the lower panels show how the
conditional SFR distributions vary with halo mass, redshift and
galaxy type. Towards lower halo masses, the distributions become broader
and less nearly log-normal, primarily because feedback has a stronger
influence on low-mass galaxies. This behaviour is not caused by unresolved
haloes, because the augmented merger trees resolve the atomic-cooling halo
population considered here.

We construct the no-SFR-scatter models by replacing the SFR entering the
reionization calculation with its median relation at fixed halo mass,
galaxy type, and redshift. The replacement is applied only to actively
star-forming galaxies; galaxies with zero SFR remain inactive. Central and
satellite galaxies are treated separately because their distinct accretion,
stripping, and feedback histories can produce systematically different
SFR--halo mass relations \citep{Mutch2016DRAGONSIII}. Combining these
populations would mix such systematic differences with the intrinsic
scatter that we aim to remove.

This deterministic construction modifies only the SFRs and corresponding
gross stellar masses mapped onto the reionization grids\footnote{Galaxy
properties continue to evolve self-consistently.
Overwriting them would also alter stellar growth, gas consumption, and
supernova feedback, preventing a controlled isolation of source
stochasticity.}. The modified radiation field can nevertheless affect later
galaxy evolution indirectly through UVB photoheating, as examined in
Section~\ref{sec:results_feedback}. We obtain the cumulative gross stellar
mass by integrating the treated SFR over time, ensuring that the cumulative
ionising budget and the SFR-weighted photoionisation rate used to estimate 
photoheating feedback are constructed from the same treated star-formation history.

As for \(f_{\rm esc}\) scatter, we construct uncalibrated and recalibrated
variants. In the uncalibrated model, replacing the scattered SFR population
with its median relation changes both the distribution of ionising sources
and their total escaped photon budget. The median replacement suppresses
the high-SFR tail and generally lowers the emissivity. In the recalibrated
model, source contributions are rescaled within each halo-mass, galaxy-type,
and redshift bin to recover the escaped source budget of the corresponding
baseline population. Comparing this model with the baseline isolates the
effect of removing SFR variations at fixed halo mass after the direct
mean-budget shift has been removed.

\subsection{X-ray stochasticity}
\label{sec:xray_stochasticity}

X-rays emitted by early galaxies heat the predominantly neutral IGM and
thereby alter the 21-cm signal through the hydrogen spin temperature. Once
Ly\(\alpha\) coupling becomes effective, increasing the gas temperature
weakens 21-cm absorption and can eventually drive the signal into emission,
while spatially inhomogeneous heating generates additional 21-cm
fluctuations \citep{PritchardFurlanetto2007,Pacucci2014XraySpectra}.
Galaxy-to-galaxy variations in X-ray luminosity can therefore affect both
the global 21-cm signal and its power spectrum.

In the standard \textsc{Meraxes} spin-temperature calculation, the
X-ray luminosity assigned to each galaxy is determined from its
star-formation activity through the adopted \(L_{\rm X}\)--SFR
relation \citep{Balu2023ThermalReionisationHistory}. This deterministic
prescription does not capture the substantial galaxy-to-galaxy
variation expected from differences in metallicity and the discrete
sampling of the high-mass X-ray binary population
\citep{Mineo2012HMXB,Lehmer2021HMXBLF}. To include this variation,
we draw the X-ray luminosity of each emitting galaxy from a log-normal
distribution about its baseline value. We adopt a scatter of
\(0.5\) dex, treat the baseline \(L_{\rm X}\)--SFR relation as the
median of the distribution, and assume that the scatter is independent
of halo mass and redshift. As for \(f_{\rm esc}\), defining a log-normal
distribution about the median increases its arithmetic mean before any
recalibration.

The scatter is applied after the galaxy properties have been evolved and
only when constructing the X-ray source field. It therefore changes neither
the SFRs and stellar masses stored by the galaxy-formation model nor the
positions of the X-ray sources. We also hold the spectral shape and energy
range of the galactic X-ray emission fixed, such that only the luminosity
normalisation of each source is modified. 

Following the treatment of the ionising sources, we construct both
uncalibrated and recalibrated X-ray scatter models. In the uncalibrated
case, the scattered galaxy luminosities are used directly to construct
the X-ray source field. This retains both the redistribution of X-ray
emission among galaxies and the accompanying change in the mean X-ray
luminosity density, and hence measures their combined effect on the thermal
and ionisation histories.

\begin{table*}
    \centering
    \caption{
    Summary of the model suite. Each of the three baseline ionising source families is
    combined with every ionising-source treatment, both with and without UVB
    feedback, giving 30 UV models (upper branch). Only the fiducial source
    family is used for the five calculations with the full spin-temperature
    field and UVB feedback (lower branch). Recalibration recovers the
    corresponding mean source emissivities; for the X-ray no-SFR-scatter
    model, both the mean X-ray and stellar Ly\(\alpha\) emissivities are
    restored, as described in
    Sections~\ref{sec:fesc}--\ref{sec:xray_stochasticity}.
    }
    \label{tab:model_variants}

    \begin{tikzpicture}[
        table node/.style={inner sep=0pt, outer sep=0pt},
        model arrow/.style={-{Stealth[length=2mm]}, line width=0.5pt}
    ]
        \node[table node] (families) {
            \footnotesize
            \setlength{\tabcolsep}{3pt}
            \begin{tabular}{@{}lc@{}}
                \toprule
                Baseline ionising source family & \(\alpha_{\rm esc}\) \\
                \midrule
                Fiducial & 0 \\
                High-mass dominated & 0.5 \\
                Low-mass dominated & -0.5 \\
                \bottomrule
            \end{tabular}
        };

        \node[table node, anchor=west] (ionising)
            at ([xshift=18mm,yshift=18mm]families.east) {
            \footnotesize
            \setlength{\tabcolsep}{3pt}
            \begin{tabular}{@{}lccc@{}}
                \toprule
                Ionising-source treatment
                & \(f_{\rm esc}\) scatter
                & SFR--halo scatter
                & Recalibrated \\
                \midrule
                Baseline
                & None & Emergent & -- \\
                \(f_{\rm esc}\) scatter
                & \(0.3\) dex & Emergent & No \\
                \(f_{\rm esc}\) scatter, recalibrated
                & \(0.3\) dex & Emergent & Yes \\
                No SFR scatter
                & None & Removed & No \\
                No SFR scatter, recalibrated
                & None & Removed & Yes \\
                \bottomrule
            \end{tabular}
        };

        \node[table node, anchor=west] (xray)
            at ([xshift=18mm,yshift=-18mm]families.east) {
            \footnotesize
            \setlength{\tabcolsep}{3pt}
            \begin{tabular}{@{}lccc@{}}
                \toprule
                X-ray source treatment
                & \(L_{\rm X}\) scatter
                & SFR--halo scatter
                & Mean(s) restored \\
                \midrule
                X-ray baseline
                & None & Emergent & -- \\
                X-ray luminosity scatter
                & \(0.5\) dex & Emergent & None \\
                X-ray luminosity scatter, recalibrated
                & \(0.5\) dex & Emergent & \(L_{\rm X}\) \\
                No SFR scatter
                & None & Removed & None \\
                No SFR scatter, recalibrated
                & None & Removed & \(L_{\rm X}\), Ly\(\alpha\) \\
                \bottomrule
            \end{tabular}
        };

        \draw[model arrow]
            ([yshift=6mm]families.east)
            -- node[above, sloped, font=\scriptsize] {all three families}
            (ionising.west);
        \draw[model arrow]
            ([yshift=2mm]families.east)
            -- node[below, sloped, font=\scriptsize] {fiducial only}
            (xray.west);
    \end{tikzpicture}
\end{table*}

For the recalibrated model, the X-ray source field is rescaled at each 
snapshot to restore the mean X-ray
luminosity density of the corresponding baseline. This preserves the
relative luminosity variations among galaxies and the associated
spatial redistribution of X-ray emission. Comparing this model with
the baseline isolates the residual effect of spatially redistributing
X-ray heating at fixed mean luminosity history. As these calculations generate spatially varying kinetic and
spin temperatures, we evaluate them
using the full spin-temperature field described in
Section~\ref{sec:powerspectrum}, rather than the saturated spin-temperature
approximation used for the main UV-ionising source suite.

On the other hand, because both \(L_{\rm X}\) and the stellar Ly\(\alpha\) emissivity are tied
to the SFR, removing scatter in the SFR--halo mass relation provides a
complementary modification to the coupling and heating source fields. We
construct uncalibrated and recalibrated
no-SFR-scatter variants using the median SFR prescription described in
Section~\ref{sec:shmr_scatter}. In the uncalibrated model, the treated SFRs
are propagated through the baseline source prescriptions, changing both the
means and spatial distributions of the X-ray and stellar Ly\(\alpha\)
emissivities. For the recalibrated variant, we rescale both coupling and heating source fields
separately at each snapshot to restore their respective baseline means.
Comparing this model with the baseline isolates the residual effect of
removing SFR variations on the spatial coupling and heating fields at fixed
mean emissivity histories.

\subsection{Model variants}
\label{sec:model_variants}

Our main model suite combines the fiducial, high-mass-dominated, and
low-mass-dominated escape-fraction families, corresponding to
\(\alpha_{\rm esc}=0\), \(0.5\), and \(-0.5\), respectively, with five
ionising-source treatments. These comprise the baseline, uncalibrated and
recalibrated \(f_{\rm esc}\)-scatter models, and uncalibrated and
recalibrated no-SFR-scatter models. The resulting 15 source configurations
are each calculated with and without UVB photoheating feedback, giving 30
UV-ionising-source models. Each comparison uses the baseline with the same
source family and feedback setting.

We examine the thermal response separately using five
fiducial calculations with the full spin-temperature field: the X-ray
baseline, uncalibrated and recalibrated models with \(0.5\)-dex scatter in
\(L_{\rm X}\) at fixed SFR, and uncalibrated and recalibrated
no-SFR-scatter models. The complete suite
therefore contains 35 models, summarised in
Table~\ref{tab:model_variants}.

\subsection{Reionization diagnostics}
\label{sec:analysis_diagnostics}

We characterise the global evolution of reionization using the
volume-averaged neutral fraction, $\langle x_{\rm HI}\rangle_{\rm V}$,
and the escaped ionising emissivity, \(\dot n_{\rm ion}\):
\begin{equation}
    \dot n_{\rm ion}(t)
    =
    \frac{N_\gamma}{V m_{\rm p}}
    \sum_{i=1}^{N_{\rm gal}}
    f_{{\rm esc},i}(t)\,{\rm SFR}_i(t),
    \label{eq:ionising_emissivity_sum}
\end{equation}
where the sum includes all \(N_{\rm gal}\) galaxies in the comoving simulation
volume \(V\), \(m_{\rm p}\) is the proton mass, and
\({\rm SFR}_i\) and \(f_{{\rm esc},i}\) are the star-formation rate and
escape fraction of galaxy \(i\), respectively. We adopt \(N_\gamma=6000\)
ionising photons per stellar baryon.

To quantify the morphology of the ionised field, we use the random-ray
mean-free-path (MFP) statistic introduced by
\citet{MesingerFurlanetto2007}. We classify cells satisfying
\(x_{\rm HI}<0.5\) as ionised. For each ray, we select an ionised cell
uniformly, draw a random point within it, and launch a ray in an
isotropically distributed direction. The ray is traced through the
periodic grid until it first enters a neutral cell, and the distance
to this phase transition is recorded as the ionised path length, \(R\). We
repeat this procedure \(N_{\rm ray}=10^{7}\) times for each neutral-fraction
grid. Because all grid cells have equal volume, this sampling produces a
volume-weighted distribution of ionised path lengths, which we use to
characterise ionised-region sizes.

\subsection{The 21-cm power spectrum}
\label{sec:powerspectrum}

In the optically thin limit and in the absence of peculiar-velocity
corrections, the differential 21-cm brightness temperature relative to the
CMB can be written as
\begin{equation}
    \delta T_{\rm b}(\boldsymbol{x},z)
    =
    T_0(z)\,
    x_{\rm HI}(\boldsymbol{x},z)
    \left[1+\delta_{\rm b}(\boldsymbol{x},z)\right]
    \left[
        1-\frac{T_{\rm CMB}(z)}
        {T_{\rm S}(\boldsymbol{x},z)}
    \right],
    \label{eq:brightness_temperature}
\end{equation}
where \(T_0(z)\approx27\sqrt{(1+z)/10}\,{\rm mK}\), \(\delta_{\rm b}\) is
the baryon density contrast, and \(T_{\rm S}\) is the hydrogen spin temperature
\citep[e.g.][]{PritchardFurlanetto2007}.

We construct the brightness-temperature field differently for the two
model suites. For the main UV-ionising source suite, we evaluate
Equation~\eqref{eq:brightness_temperature} in the saturated
spin-temperature limit, \(T_{\rm S}\gg T_{\rm CMB}\), and neglect
peculiar-velocity gradients. As all models use the same density field supplied by
the \textit{N}-body simulation, differences among their 21-cm fields then
arise only from the neutral-fraction distribution.
For the separate X-ray-stochasticity calculations, we retain the
spatially varying spin-temperature field computed by
\textsc{Meraxes} following the implementation of
\citet{Balu2023ThermalReionisationHistory}, and use the brightness-temperature
field computed directly by \textsc{Meraxes}. Unlike the approximation in
Equation~\eqref{eq:brightness_temperature}, this field includes finite
optical depth and the local line-of-sight peculiar-velocity-gradient
correction. 

We quantify the fluctuations using the dimensionless, spherically averaged
21-cm power spectrum,
\begin{equation}
    \Delta_{21}^{2}(k,z)
    =
    \frac{k^{3}}{2\pi^{2}}P_{21}(k,z),
    \label{eq:21cm_power_spectrum}
\end{equation}
where \(P_{21}(k,z)\) is the power spectrum of
\(\delta T_{\rm b}(\boldsymbol{x},z)\), and \(\Delta_{21}^{2}\) is
expressed in \({\rm mK}^{2}\). Our primary comparison uses the bin centred
at \(k_{\rm eff}\simeq1\,{\rm cMpc}^{-1}\), which probes relatively
small-scale structure in the ionisation field and, for the X-ray models,
the thermal field.

We estimate the power-spectrum uncertainty using
\textsc{21cmSense} v2 \citep{Murray2024}. We adopt the
{\it SKA}-Low AA4 configuration and assume observations of a
single tracked deep field for \(180\) days at \(6\,{\rm h}\) per day,
giving a total integration time of \(1080\,{\rm h}\), with an
\(8\,{\rm MHz}\) bandwidth. Baselines longer than \(1\,{\rm km}\) are
excluded. We adopt the moderate foreground prescription, which
removes foreground-contaminated modes within the wedge and applies a
\(0.1\,h\,{\rm Mpc}^{-1}\) buffer beyond the horizon limit. The
resulting total \(1\sigma\) uncertainty includes both instrumental
thermal noise and sample variance. For each model family,
the full power spectrum of its baseline model is used when calculating
the sample-variance contribution.

\section{Impact of source stochasticity on reionization and the 21-cm signal}
\label{sec:results}

We now test whether galaxy-to-galaxy stochasticity remains important once
its effect on the mean radiative emissivity is removed. The uncalibrated
models show the combined response to changing the mean and redistributing
radiation, whereas the recalibrated models isolate the latter. This
comparison determines whether stochasticity itself remains a major driver
after the population mean has been controlled. We first examine the
ionising emissivity, global reionization history, and ionisation morphology,
before considering the 21-cm signatures of ionising and X-ray sources.
Unless otherwise stated, all models include UVB photoheating feedback.

\subsection{Ionising emissivity and the global reionization history}
\label{sec:results_global}

\begin{figure*}
    \centering
    \includegraphics[width=0.84\linewidth]
    {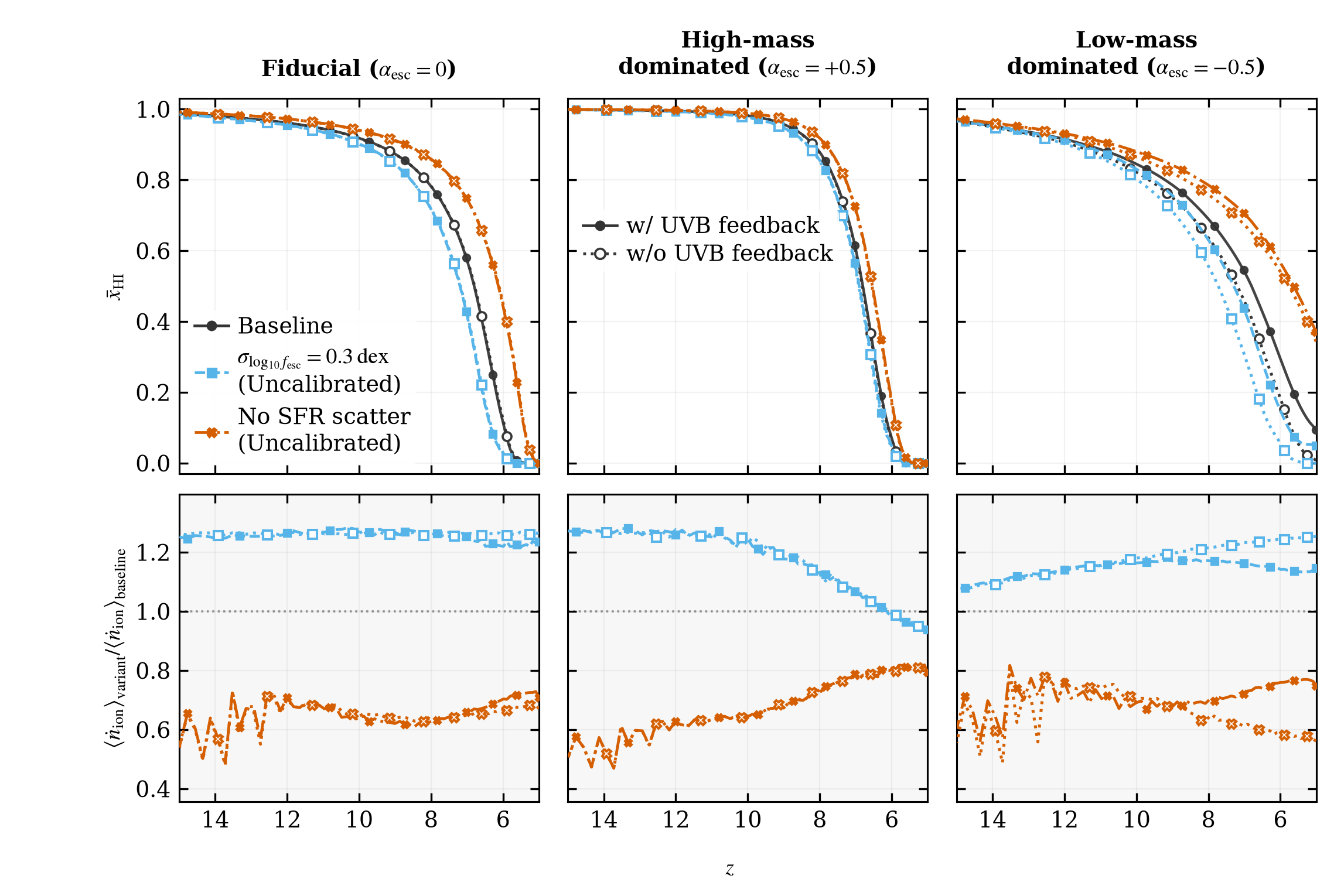}
    \caption{ \label{fig:global_uncalibrated}
    Global reionization histories and escaped ionising-emissivity ratios for
    the uncalibrated source modifications. The columns show the fiducial
    (\(\alpha_{\rm esc}=0\)), high-mass-dominated
    (\(\alpha_{\rm esc}=0.5\)), and low-mass-dominated
    (\(\alpha_{\rm esc}=-0.5\)) source families. The upper panels show the
    volume-averaged neutral fraction. The lower panels show the emissivity
    of each modified variant relative to the corresponding baseline with
    the same UVB-feedback setting. Grey circles denote the baseline models,
    blue squares denote models with \(0.3\)-dex scatter in \(f_{\rm esc}\),
    and orange crosses denote the no-SFR-scatter models. Filled and open
    symbols distinguish calculations with and without UVB feedback,
    respectively.
    }
    \includegraphics[width=0.84\linewidth]
    {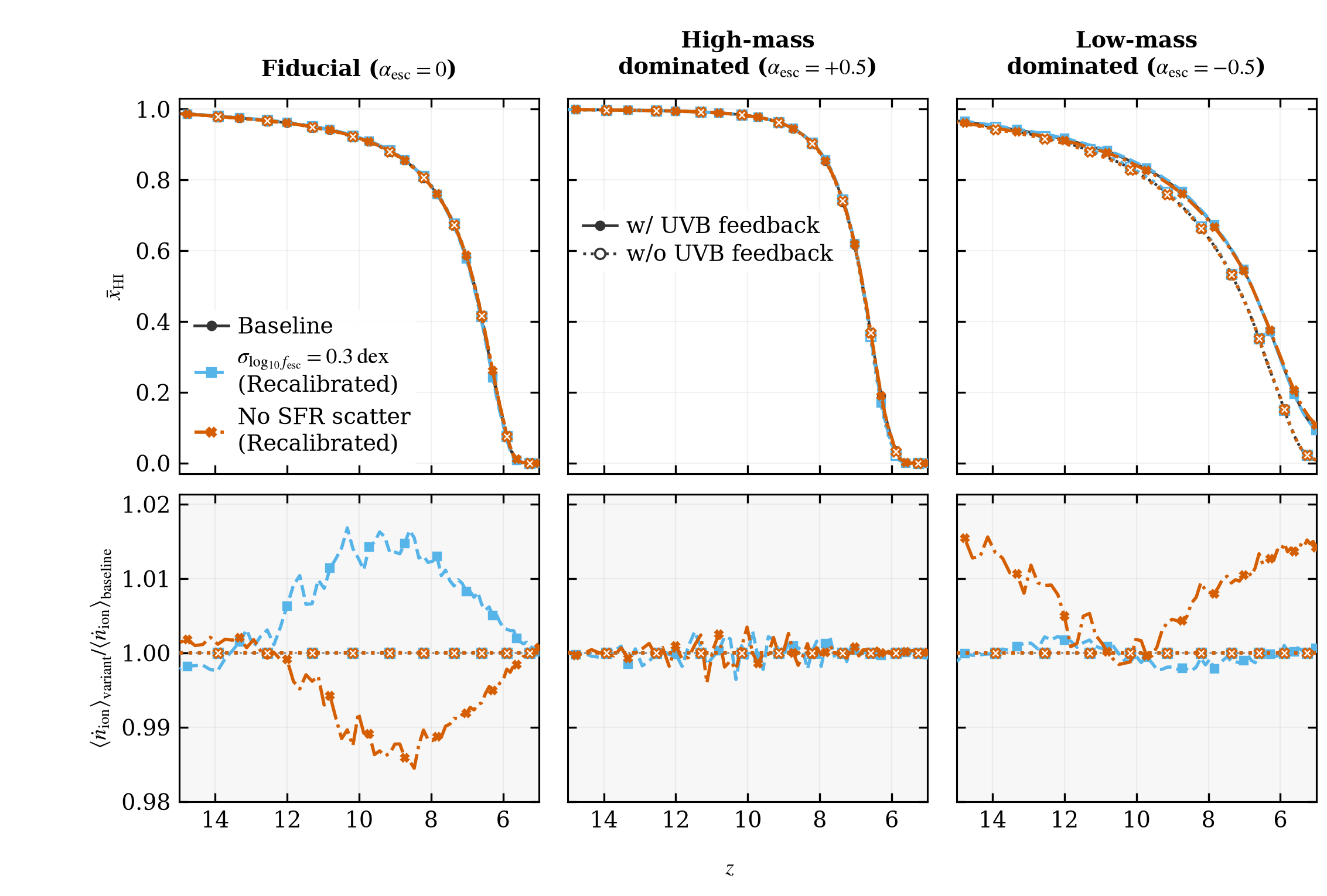}
    \caption{\label{fig:global_recalibrated}Same as Fig.~\ref{fig:global_uncalibrated}, but for the recalibrated
    source modifications. The differences from the corresponding baselines
    are substantially reduced after matching the mean escaped ionising
    emissivity.}
\end{figure*}

Figure~\ref{fig:global_uncalibrated} compares the global neutral fraction
and escaped ionising emissivity across the three source families.
The three baseline models with UVB feedback are calibrated to a common
midpoint, reaching \(\langle x_{\rm HI}\rangle_{\rm V}=0.5\) at
\(z\simeq6.80\). They differ substantially in duration. Defining
\(\Delta z_{\rm EoR}=z_{75}-z_{25}\), where \(z_{75}\) and \(z_{25}\)
correspond to \(\langle x_{\rm HI}\rangle_{\rm V}=0.75\) and \(0.25\),
respectively, we find \(\Delta z_{\rm EoR}=1.5\), \(1.0\), and \(2.7\)
for the fiducial, high-mass-dominated, and low-mass-dominated models.
These different ionisation histories yield CMB Thomson scattering optical
depths of \(\tau_{\rm e}=0.0584\), \(0.0510\), and \(0.0654\),
respectively. Weighting the escaped emissivity towards rarer, more
strongly clustered massive haloes produces a comparatively rapid
transition. The larger abundance of low-mass galaxies extends the
ionising contribution over a broader redshift interval
\citep{McQuinn2007Morphology,Geil2016Morphology}.

The lower panels of Fig.~\ref{fig:global_uncalibrated} show that both
modifications first act through the mean escaped emissivity. Adding
\(0.3\)-dex scatter in \(f_{\rm esc}\) generally increases the emissivity,
whereas replacing the scattered SFR--halo mass relation with its median
reduces it. These responses follow from the linear source weighting in
Equation~\eqref{eq:intro_ionising_budget}: upward fluctuations in a
log-normal \(f_{\rm esc}\) distribution raise its arithmetic mean in the
absence of upper-bound saturation, while removing the high-SFR tail lowers
the mean ionising contribution.

The emissivity change depends on which halo masses supply most
of the ionising photons. For the fiducial family, \(f_{\rm esc}\) scatter increases
the emissivity by approximately \(25\) per cent over
\(5\lesssim z\lesssim15\), with little dependence on UVB feedback.
In the high-mass-dominated family, the emissivity ratio decreases from
approximately \(1.27\) at high redshift to unity near \(z\simeq6\),
then falls slightly below the baseline towards \(z\simeq5\).
An increasing fraction of the dominant sources approach the physical
upper bound \(f_{\rm esc}=1\). This bound clips their upward fluctuations,
and downward fluctuations can reduce their mean escaped contribution.
The low-mass-dominated family shows a growing enhancement towards lower
redshift in the calculation without UVB feedback, reaching approximately
\(25\) per cent near \(z\simeq5\). Including UVB feedback moderates this
growth, giving an enhancement of approximately \(15\) per cent at the
same redshift.

Removing SFR scatter produces a sustained reduction in the escaped
emissivity. The fiducial no-SFR-scatter model remains approximately
\(30\)--\(40\) per cent below its baseline over much of the redshift
range shown. In the high-mass-dominated family, the emissivity rises
from roughly half of the baseline value at high redshift to about
\(80\) per cent near \(z\simeq5\). The corresponding calculations with
and without UVB feedback remain close in both families. The
low-mass-dominated model develops a stronger dependence on feedback:
near \(z\simeq5\), the no-SFR-scatter emissivity is approximately
\(75\) per cent of the baseline value with UVB feedback and about
\(60\) per cent without it. The redshift dependence reflects the
evolving width and shape of the SFR distribution, together with the halo
masses that dominate the ionising emissivity
(see Fig.~\ref{fig:source_relations}) and their response to photoheating.

These emissivity shifts propagate directly into the global
reionization histories. Around the common baseline midpoint at
\(z\simeq6.8\), introducing \(f_{\rm esc}\) scatter lowers
\(\langle x_{\rm HI}\rangle_{\rm V}\) to approximately \(0.32\),
\(0.44\), and \(0.38\) in the fiducial, high-mass-dominated, and
low-mass-dominated families, respectively. Removing SFR scatter raises
the neutral fractions to approximately \(0.71\), \(0.64\), and
\(0.68\). Thus, the uncalibrated modifications can shift the neutral
fraction by approximately \(\Delta\langle x_{\rm HI}\rangle_{\rm V}
\simeq0.2\) around the baseline midpoint. The low-mass-dominated
no-SFR-scatter model retains a neutral fraction of approximately
\(0.37\) at the final output, \(z\simeq5.01\).

\begin{figure*}
    \centering
    \includegraphics[width=0.95\linewidth]
    {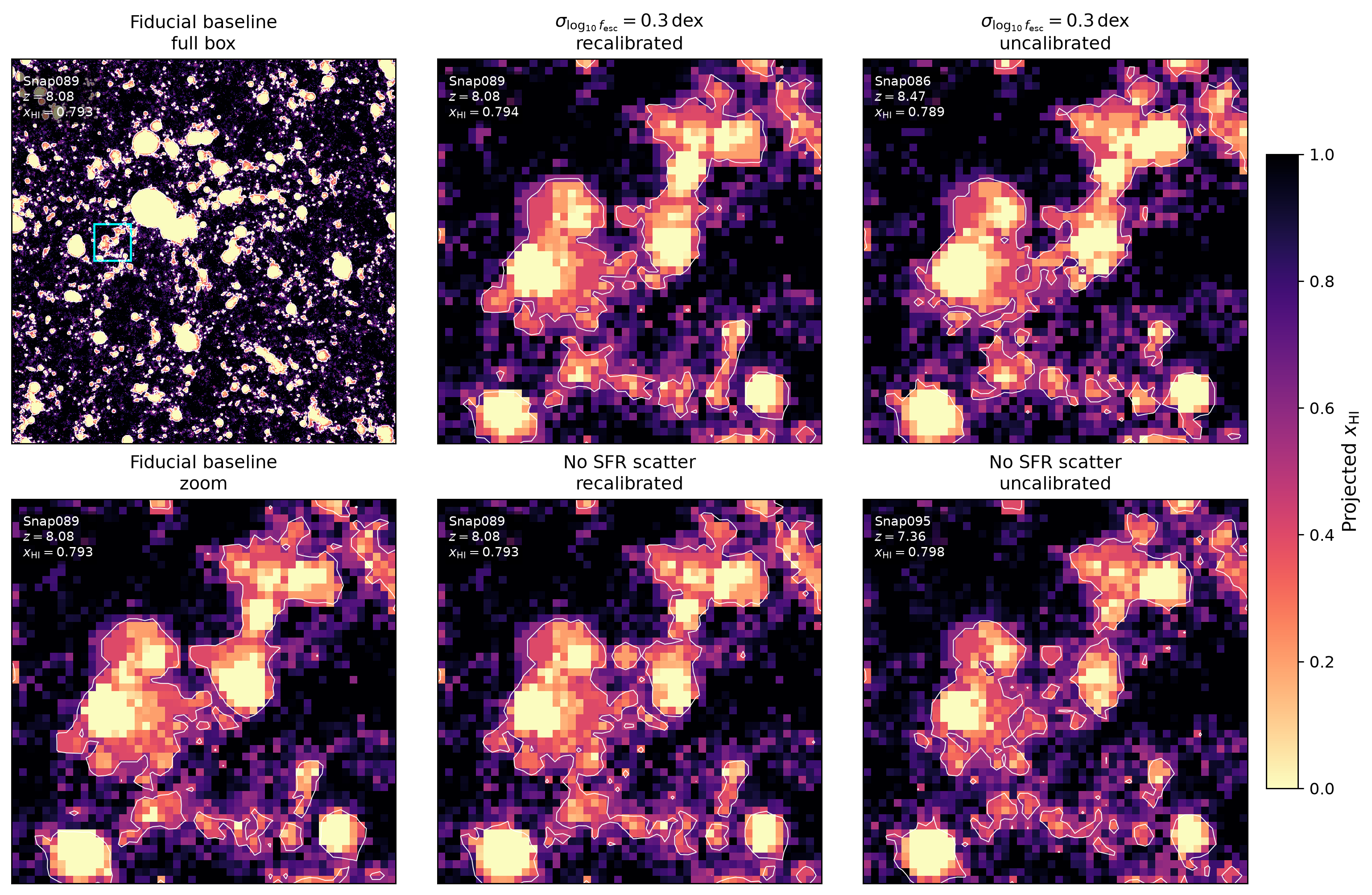}
    \caption{
    Projected neutral-fraction fields with a projection depth of
    approximately \(3\,{\rm cMpc}\) for the fiducial source family at
    \(\langle x_{\rm HI}\rangle_{\rm V}\simeq0.79\).
    The upper-left panel shows the full baseline volume, with the
    highlighted region enlarged in the remaining panels. These compare
    the baseline with the recalibrated and uncalibrated \(0.3\)-dex
    \(f_{\rm esc}\)-scatter and no-SFR-scatter models.
    Each panel lists its snapshot, redshift, and volume-averaged
    neutral fraction.
    }
    \label{fig:ionisation_slice}
\end{figure*}

The calculations without UVB feedback retain the same ordering of the
source modifications. Their global histories remain close to the
corresponding calculations with feedback in the fiducial and
high-mass-dominated families. The low-mass-dominated family responds more
strongly: disabling photoheating advances the baseline midpoint from
\(z\simeq6.80\) to \(7.20\), with the modified models also reaching
their midpoints at higher redshift.
This dependence is consistent with the greater contribution of
low-mass galaxies which can be more easily affected by the changes in the radiation field.
The uncalibrated calculations therefore combine changes in the mean escaped
emissivity with source redistribution and the response of galaxy evolution
to the resulting radiation field. Their large shifts in the reionization
history therefore do not by themselves demonstrate a strong independent
effect of galaxy-to-galaxy stochasticity. We next use the recalibrated
models to separate these contributions.

\subsection{Separating mean-emissivity and redistribution effects}
\label{sec:results_recalibrated_global}

Figure~\ref{fig:global_recalibrated} shows the global reionization
histories and escaped ionising emissivities after recalibrating the source
modifications to remove their direct effect on the mean emissivity.
The recalibrated neutral fraction histories are nearly indistinguishable
from their corresponding baselines in all three source families.
This agreement holds for the calculations with and without UVB
feedback, with shifts in the reionization midpoint smaller than
\(\Delta z\simeq0.02\). The convergence shows that the changes in the
uncalibrated reionization histories are driven predominantly by changes
in the mean escaped emissivity. This is the central distinction: once that
mean is controlled, altering the galaxy-to-galaxy distribution of the
ionising output has little effect on the global history.

Small emissivity differences develop during the subsequent coupled
evolution. In the fiducial family, the \(f_{\rm esc}\)-scatter and
no-SFR-scatter models show residuals of opposite sign, with amplitudes
below approximately \(2\) per cent. The high-mass-dominated models
remain very close to their baseline. In the low-mass-dominated family,
the clearest response occurs in the no-SFR-scatter model, whose
emissivity is enhanced by up to approximately \(1.5\) per cent.

The calculations without UVB feedback identify the origin of these
residuals. Their recalibrated emissivities remain close to the baseline
values throughout the redshift range shown, and their global
reionization histories nearly coincide within each source family.
With feedback included, the modified source distributions produce
different ionisation and UVB fields, which can alter the subsequent gas
supply and star formation of low-mass galaxies. The resulting populations
therefore develop small differences in their escaped emissivity. This
response is clearest in the low-mass-dominated family, where a greater
fraction of the emissivity is supplied by galaxies susceptible to
photoheating.

Nevertheless, the weak response of the global neutral fraction shows that these
feedback-generated emissivity differences have little effect on the
overall timing of reionization. Recalibration thus separates the dominant
mean-emissivity effect from the weaker response to source redistribution
and UVB feedback. We next ask whether this residual response changes the
morphology of the ionised field.

\subsection{Ionisation morphology}
\label{sec:results_feedback}

We compare the ionisation morphology at fixed
\(\langle x_{\rm HI}\rangle_{\rm V}\), thereby controlling for the global
stage of reionization\footnote{Because the recalibrated histories within
each source family are very similar, matching
$\langle x_{\rm HI}\rangle_{\rm V}$ generally also selects snapshots at
similar redshifts. Different source families can nevertheless reach the
same neutral fraction at different redshifts because their global
reionization histories differ.}. The recalibrated models then isolate the
morphological response to source redistribution and its coupling to UVB
feedback.

\begin{figure*}
    \centering
    \includegraphics[width=0.95\linewidth]
    {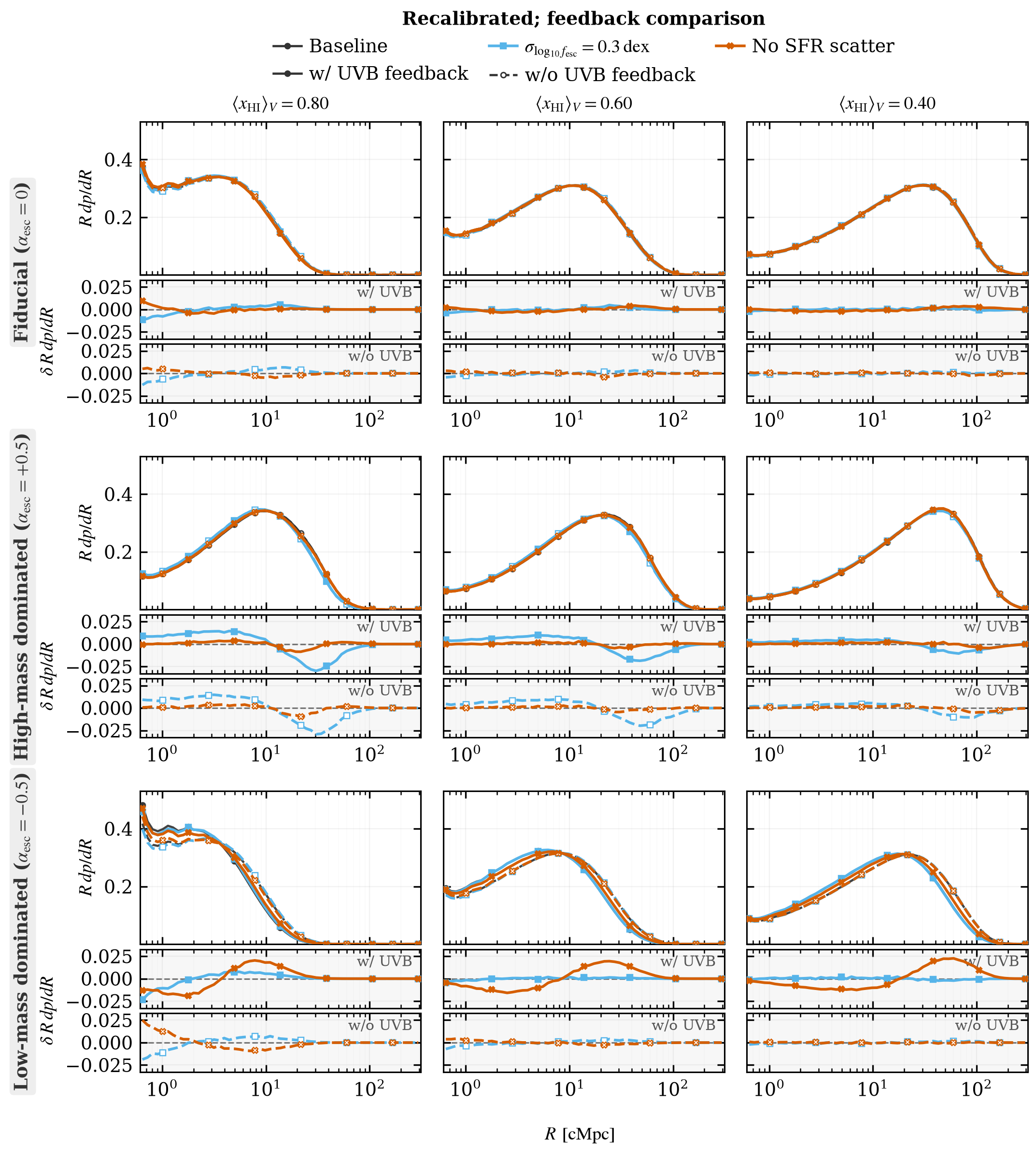}
   \vspace*{-3mm} 
   \caption{
    Ionised-region random-ray MFP distributions for the recalibrated
    models at \(\langle x_{\rm HI}\rangle_{\rm V}=0.8\), \(0.6\), and
    \(0.4\) (columns). The three groups of rows show the fiducial,
    high-mass-dominated, and low-mass-dominated source families.
    Grey circles denote the baselines, blue squares the \(0.3\)-dex
    \(f_{\rm esc}\)-scatter models, and orange crosses the
    no-SFR-scatter models. Filled and open symbols denote calculations
    with and without UVB feedback, respectively. Within each group,
    the upper panels show \(R\,{\rm d}p/{\rm d}R\), and the two smaller
    panels show differences \(\delta R\,{\rm d}p/{\rm d}R\) from the
    corresponding baseline with the same feedback treatment, with
    feedback above and without feedback below. Ionised cells satisfy
    \(x_{\rm HI}<0.5\).
    }
    \label{fig:mfp_recalibrated}
\end{figure*}

Figure~\ref{fig:ionisation_slice} provides a visual comparison of the
fiducial source family at an early stage of reionization,
$\langle x_{\rm HI}\rangle_{\rm V}\simeq0.79$\footnote{The uncalibrated
models are shown for illustration only. Because they reach the same neutral
fraction at different redshifts, their spatial differences include both the
modified source prescription and the additional cosmic evolution required
to reach that stage. Our quantitative morphology comparison therefore
focuses on the recalibrated models.}. After matching the mean escaped
emissivity, the ionisation fields remain visually similar to the baseline.
The remaining differences are subtle, suggesting that source
redistribution has only a weak effect on the large-scale morphology of the
fiducial model.

\begin{figure*}
    \centering
    \includegraphics[width=0.95\linewidth]{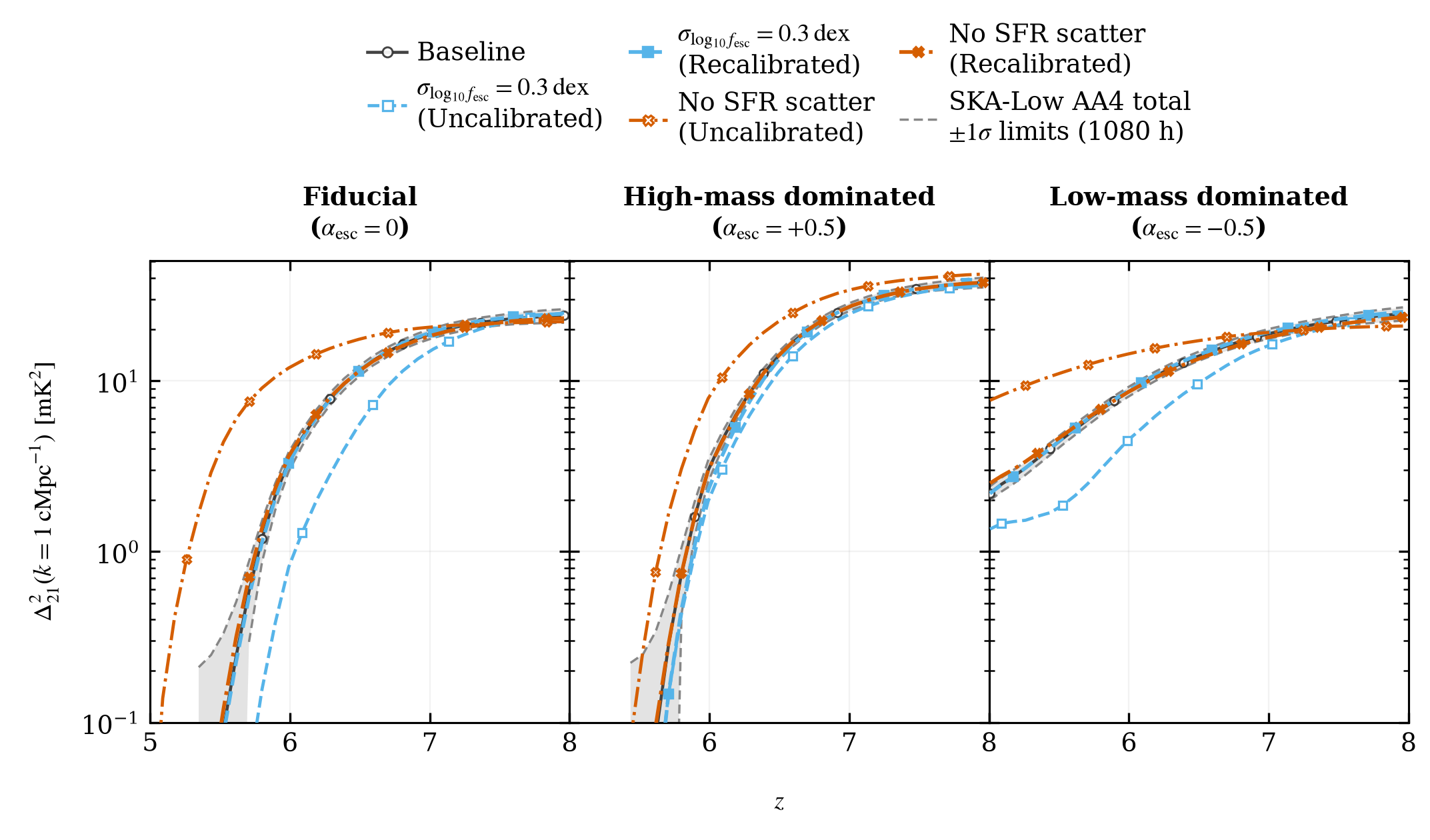}
\vspace*{-3mm}
    \caption{
    Redshift evolution of the 21-cm power at
    \(k\simeq1\,{\rm cMpc}^{-1}\)
    (\(k_{\rm eff}=1.005\,{\rm cMpc}^{-1}\)) for the fiducial
    (\(\alpha_{\rm esc}=0\)), high-mass-dominated
    (\(\alpha_{\rm esc}=0.5\)), and low-mass-dominated
    (\(\alpha_{\rm esc}=-0.5\)) source families.
    All models include UVB feedback.
    Black circles show the family baselines. Blue squares show the
    \(0.3\)-dex \(f_{\rm esc}\)-scatter models, and orange crosses
    show the no-SFR-scatter models. Open symbols denote uncalibrated
    models and filled symbols denote recalibrated models.
    The brightness-temperature field is evaluated in the saturated
    spin-temperature limit using the neutral-fraction and
    matter-density fields, without redshift-space distortions.
    The grey shaded regions, bounded by dashed lines, show the total
    \(1\sigma\) {\it SKA}-Low AA4 uncertainty around each family baseline,
    including thermal noise and sample variance, for a \(1080\)-h
    observation.
    }
    \label{fig:power_k1}
\end{figure*}

Figure~\ref{fig:mfp_recalibrated} quantifies the residual morphology using
random-ray MFP distributions with and without UVB feedback. The
corresponding uncalibrated distributions are shown for completeness in
Appendix~\ref{app:uncalibrated_morphology}; these combine source
redistribution with changes in the mean emissivity and cosmic evolution.
We therefore focus on the recalibrated models, for which changes in the
distributions trace residual morphological differences rather than shifts
in the global reionization history. In the fiducial family, the MFP
distributions remain close to the baseline at all three stages. At
\(\langle x_{\rm HI}\rangle_{\rm V}=0.8\), adding \(f_{\rm esc}\) scatter
weakly shifts the distribution towards longer paths, whereas removing SFR
scatter shifts it towards shorter paths. These trends are consistent with
changes in the contrast of the ionising source field, but remain small
compared with the evolution of the distribution during reionization.

The high-mass-dominated \(f_{\rm esc}\)-scatter model instead shifts
towards shorter paths, most clearly at
\(\langle x_{\rm HI}\rangle_{\rm V}=0.8\). This response is consistent
with the escape-fraction saturation discussed in
Section~\ref{sec:results_global}. The massive sources associated with the
largest ionised regions already have escape fractions close to unity, so
clipping suppresses their upward fluctuations while retaining the downward
ones. Recalibration restores the mean emissivity but preserves this change
in the relative weighting of sources. The same shift towards shorter paths
appears without UVB feedback, supporting an origin in the bounded
escape-fraction prescription rather than feedback.

The largest feedback-dependent response occurs in the low-mass-dominated
no-SFR-scatter model. With UVB feedback, this model develops
systematically longer paths at all three neutral fractions; without
feedback, it shows only a weak shift towards shorter paths. The reversal
demonstrates that instantaneous source redistribution alone does not
produce the longer paths. Instead, they arise from its coupling to UVB
feedback, consistent with the sensitivity of low-mass galaxies to
photoheating and with the feedback-dependent emissivity residuals in
Fig.~\ref{fig:global_recalibrated}. The detailed source-by-source pathway
is not isolated here, but the comparison shows that the residual morphology
depends on both the halo masses supplying the escaped emissivity and their
response to photoheating. Even in this most responsive case, the
morphological effect remains secondary to the change in reionization timing
produced when the mean emissivity is allowed to shift.

\subsection{Impact on the 21-cm power spectrum}
\label{sec:results_21cm}

\begin{table}
    \centering
    \caption{
    Maximum deviations of the with-UVB-feedback models from their
    corresponding family baselines at \(k=1\,{\rm cMpc}^{-1}\)
    over \(5\leq z\leq8\).
    We define
    \(D=(\Delta^2_{21,\rm variant}-\Delta^2_{21,\rm baseline})
    /\sigma_{\rm SKA}\),
    where \(\sigma_{\rm SKA}\) is the forecast total \(1\sigma\)
    uncertainty for a \(1080\)-h {\it SKA}-Low AA4 observation of the
    corresponding baseline, and \(D_{\rm peak}\) is the signed deviation at the
    redshift \(z_{\rm peak}\) where \(|D|\) is largest.
    Positive and negative values indicate power above and below
    the baseline, respectively.
    }
    \label{tab:power_k1_sigma}
    \begin{tabular}{llrrrr}
        \toprule
        Source & \multirow{2}{*}{Modification}
        & \multicolumn{2}{c}{Uncalibrated}
        & \multicolumn{2}{c}{Recalibrated} \\
        \cmidrule(lr){3-4}
        \cmidrule(lr){5-6}
        family & & \(D_{\rm peak}\) & \(z_{\rm peak}\)
          & \(D_{\rm peak}\) & \(z_{\rm peak}\) \\
        \midrule
        \multirow{2}{*}{Fiducial}
        & \(0.3\)-dex \(f_{\rm esc}\) scatter
        & \(-7.74\) & \(6.29\) & \(+0.34\) & \(7.48\) \\
        & No SFR scatter
        & \(+24.27\) & \(5.80\) & \(+0.56\) & \(5.89\) \\
        \addlinespace
        High-mass
        & \(0.3\)-dex \(f_{\rm esc}\) scatter
        & \(-3.41\) & \(6.39\) & \(-1.70\) & \(6.09\) \\
        dominated
        & No SFR scatter
        & \(+12.22\) & \(6.19\) & \(-0.19\) & \(6.19\) \\
        \addlinespace
        Low-mass
        & \(0.3\)-dex \(f_{\rm esc}\) scatter
        & \(-8.01\) & \(5.80\) & \(+0.36\) & \(6.81\) \\
        dominated & No SFR scatter
        & \(+26.07\) & \(5.01\) & \(+1.48\) & \(5.01\) \\
        \bottomrule
    \end{tabular}
\end{table}

\begin{figure*}
    \centering
    \includegraphics[width=\linewidth]{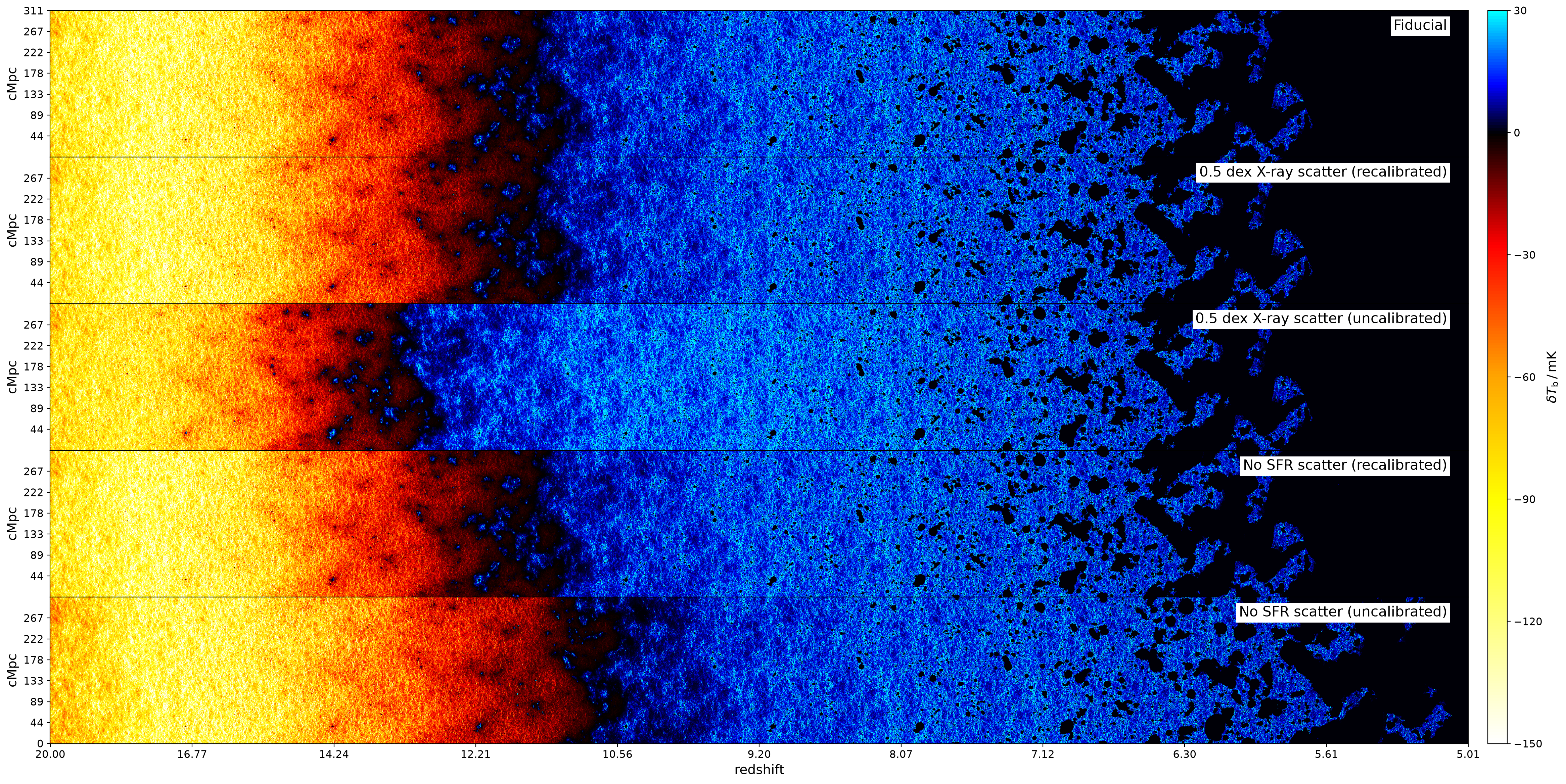}
    \caption{
    Lightcone slices of the differential 21-cm brightness temperature
    for the fiducial baseline, the recalibrated and uncalibrated
    X-ray-scatter models, and the recalibrated and
    uncalibrated no-SFR-scatter models, from top to bottom.
    The X-ray-scatter models include \(0.5\)-dex scatter in \(L_{\rm X}\)
    at fixed SFR. In the recalibrated no-SFR-scatter model, both the mean
    X-ray and stellar Ly\(\alpha\) emissivities are restored. The same
    transverse slice is shown for each model,
    with redshift decreasing from approximately \(20\) to \(5\)
    from left to right. All panels use the full spin-temperature
    field and share the same colour scale.
    }
    \label{fig:xray_lightcones}
\end{figure*}

Figure~\ref{fig:power_k1} shows how these changes propagate into the
21-cm power at \(k=1\,{\rm cMpc}^{-1}\) in the saturated
spin-temperature limit. The uncalibrated models depart strongly from their
corresponding baselines because the emissivity changes shift the global
reionization histories (Section~\ref{sec:results_global}). At a fixed
redshift, the models can therefore occupy substantially different stages
of reionization. The earlier reionization in the \(f_{\rm esc}\)-scatter
models reduces the power near the end of reionization, whereas removing
SFR scatter delays this decline and retains stronger fluctuations at the
same redshift.

Once the mean escaped emissivity is matched, these differences are strongly
suppressed in all three source families. The fiducial models closely follow
their baseline throughout the redshift range shown. The largest residuals
occur in the high-mass-dominated \(f_{\rm esc}\)-scatter model and the
low-mass-dominated no-SFR-scatter model towards lower redshift. Because
these comparisons are made at fixed redshift, the residuals include both
small offsets in reionization timing and changes in ionisation morphology.

Table~\ref{tab:power_k1_sigma} quantifies the largest departures
relative to the forecast statistical uncertainties of a \(1080\)-h
{\it SKA}-Low AA4 observation, also shown by the grey regions in
Fig.~\ref{fig:power_k1}. We define $D=(\Delta^2_{21,\rm variant}-\Delta^2_{21,\rm baseline})
/\sigma_{\rm SKA}$, where \(\sigma_{\rm SKA}\) is the forecast total \(1\sigma\)
uncertainty using the corresponding baseline models. We see the uncalibrated
\(f_{\rm esc}\)-scatter models reach maximum absolute deviations
of approximately \(7.7\sigma\), \(3.4\sigma\), and \(8.0\sigma\)
in the fiducial, high-mass-dominated, and low-mass-dominated
families, respectively. The corresponding no-SFR-scatter models
produce larger deviations of approximately \(24.3\sigma\),
\(12.2\sigma\), and \(26.1\sigma\).
After recalibration, the differences generally fall below \(1\sigma\).
The two largest residuals are \(-1.70\sigma\) for the high-mass-dominated
\(f_{\rm esc}\)-scatter model and \(+1.48\sigma\) for the
low-mass-dominated no-SFR-scatter model. Thus, every recalibrated model
remains within \(2\sigma\) of its family baseline over the redshift range
considered. The large apparent 21-cm signature of the uncalibrated models is
therefore primarily a signature of their shifted mean emissivities, rather
than of source stochasticity itself. At this scale, the remaining effect of
redistributing a fixed emissivity is unlikely to be distinguishable even
with a deep {\it SKA} observation.

\subsection{Impact of X-ray stochasticity on the 21-cm signal}
\label{sec:results_xray}

We finally consider the thermal channel using the fiducial calculations
described in Section~\ref{sec:xray_stochasticity}. These calculations follow
the full spin-temperature field and compare the baseline with uncalibrated
and recalibrated models that either introduce \(0.5\)-dex scatter in \(L_{\rm X}\) at
fixed SFR or remove the intrinsic scatter in the SFR--halo
mass relation.

Figure~\ref{fig:xray_lightcones} compares the heating histories across the
five X-ray-source calculations. The uncalibrated
X-ray-scatter model has
weaker absorption during the heating epoch and develops regions of positive
emission at higher redshift than the baseline. The average absorption
minimum correspondingly shifts from \(\sim-115\,{\rm mK}\) at \(z\sim17.5\)
in the baseline to \(\sim-95\,{\rm mK}\) at \(z\sim18.5\), while the global
signal changes from absorption to emission at \(z\sim13\), rather than
\(z\sim11.5\). These changes reflect the increased mean X-ray emissivity,
which heats the neutral IGM earlier. After recalibration, the absorption and
emission patterns are visually very similar to the baseline, and the
volume-averaged brightness temperature differs by less than
\(0.1\,{\rm mK}\) over the redshift range shown.

By contrast, the uncalibrated no-SFR-scatter model produces the opposite
shift. Replacing the scattered SFR population with its median relation
reduces the mean X-ray and stellar Ly\(\alpha\) emissivities, delaying
Ly\(\alpha\) coupling and heating. Consequently, the global signal changes
from absorption to emission at \(z\sim11\), compared with \(z\sim11.5\) in
the baseline. After matching both mean emissivities, the recalibrated model
closely follows the baseline: the transition returns to \(z\sim11.5\), and
the depth of the absorption minimum agrees to within approximately
\(0.05\) per cent.

\begin{figure}
    \begin{minipage}{\columnwidth}
    \centering
    \includegraphics[width=\linewidth]
    {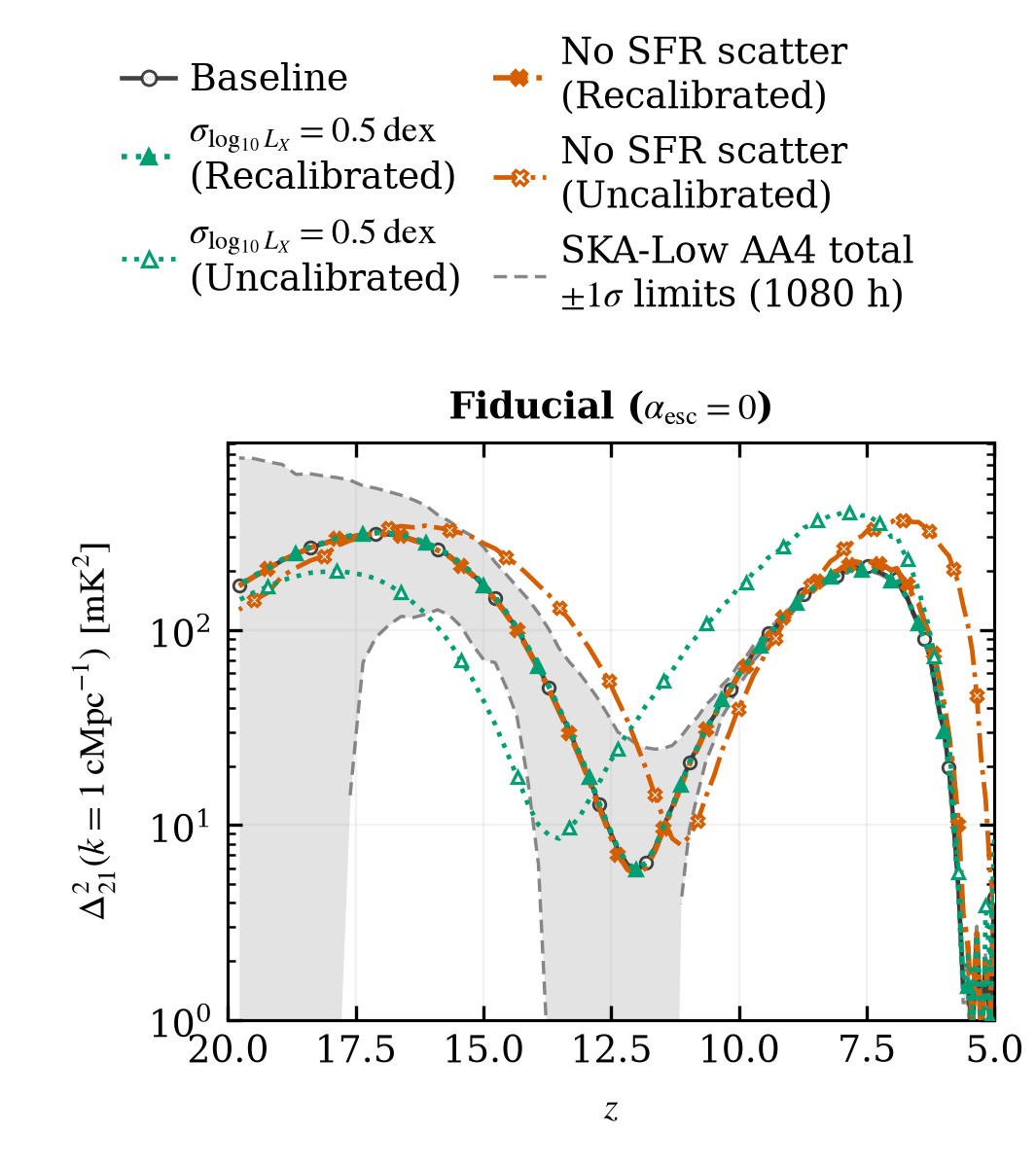}
    \caption{
    Redshift evolution of the dimensionless 21-cm power at
    \(k\simeq1\,{\rm cMpc}^{-1}\) for the fiducial
    X-ray-source calculations. Black circles show the fiducial
    baseline. Green triangles show the models with \(0.5\)-dex scatter
    in \(L_{\rm X}\) at fixed SFR, and orange crosses show the
    no-SFR-scatter models. Open and filled symbols denote
    the uncalibrated and recalibrated treatments, respectively. The
    brightness-temperature field includes the spatially varying spin
    temperature. The grey shaded region, bounded by dashed lines, shows the total
    \(1\sigma\) {\it SKA}-Low AA4 uncertainty around the baseline,
    including thermal noise and sample variance, for a \(1080\)-h
    observation.
    }
    \label{fig:xray_power_spectra}
    \end{minipage}
\end{figure}

Figure~\ref{fig:xray_power_spectra} shows the corresponding evolution
of the 21-cm power at \(k\sim1\,{\rm cMpc}^{-1}\).
The two uncalibrated source modifications shift the power
minimum in opposite directions.
For X-ray scatter, the minimum between the two broad peaks shifts from
\(z\sim12\) in the baseline to \(z\sim13.5\) in the
uncalibrated model, consistent with the earlier heating seen in
the lightcones. At a fixed redshift, this shift
can suppress or enhance the power. The uncalibrated power is
approximately \(0.17\) times the baseline value at
\(z\sim14\) and 3--4 times the baseline
at \(z\sim10.5\). The two models reach different stages of
heating at the same redshift, changing both the contrast with
the CMB and the spatial fluctuations in spin temperature.
The no-SFR-scatter model instead shifts the minimum to
\(z\sim11\), consistent with its delayed heating.

After recalibration, the \(L_{\rm X}\)-scatter model closely follows
the baseline, with power differences below \(2\) per cent over
\(10\lesssim z\lesssim20\). For the no-SFR-scatter model, matching both
mean emissivities reduces the power differences to below \(6\) per cent
over the same redshift interval, with the largest difference, approximately
\(5.6\) per cent, occurring at \(z\sim12\). The power differences
in both recalibrated models lie within the forecast \(1\sigma\)
uncertainty of the \(1080\)-h {\it SKA}-Low AA4 observation
throughout this interval.

The dominant apparent response therefore arises from changes in the
population-averaged source emissivities: the X-ray emissivity for the
\(L_{\rm X}\)-scatter model, and both the X-ray and stellar Ly\(\alpha\)
emissivities for the no-SFR-scatter model. Restoring the appropriate
baseline means brings the timing of coupling and heating, the global
absorption signal, and the power spectrum into close agreement with the
baseline.
As for the ionising sources, the remaining effect is
subdominant once the relevant population means are consistently calibrated.

\section{Conclusion}
\label{sec:conclusion}

Our main result is that galaxy-to-galaxy stochasticity is not a major driver
of the reionization, heating, and 21-cm observables considered here once the
corresponding mean emissivities are consistently calibrated. We establish
this using the
\textsc{Meraxes} semi-analytic galaxy-formation and reionization model.
Across three escape-fraction prescriptions, we introduce \(0.3\)-dex scatter
in \(f_{\rm esc}\) and separately remove the intrinsic scatter in the
SFR--halo mass relation. Comparing uncalibrated and recalibrated variants
separates changes in the mean escaped ionising emissivity from the residual
effect of redistributing that emissivity among galaxies. Separate
calculations following the full spin-temperature field introduce
\(0.5\)-dex scatter in X-ray luminosity at fixed SFR
and separately remove the intrinsic SFR scatter at fixed
halo mass. For the latter, the recalibration restores both the mean X-ray
and stellar Ly\(\alpha\) emissivities.

The dominant response of the ionising-source models is a change in the
mean escaped emissivity. Scatter in \(f_{\rm esc}\) generally increases
this emissivity and advances reionization, whereas replacing the scattered
SFR population with its median relation reduces the emissivity and delays
reionization. At the common baseline midpoint, \(z\simeq6.8\), the
uncalibrated \(f_{\rm esc}\)-scatter models have
\(\langle x_{\rm HI}\rangle_{\rm V}\simeq0.32\)--\(0.44\), while the
no-SFR-scatter models have
\(\langle x_{\rm HI}\rangle_{\rm V}\simeq0.64\)--\(0.71\), compared with
the baseline value of \(0.5\). These shifts show that stochastic source
prescriptions can substantially change the global history through their
effect on the mean emissivity. They should therefore be interpreted as a
change in the population-averaged source model, rather than as an
independent spatial imprint of stochasticity.

After matching the mean escaped emissivity, the reionization midpoints
agree with their corresponding baselines to within \(\Delta z\simeq0.02\),
and the remaining emissivity differences are below approximately \(2\) per
cent. These residuals largely disappear when UVB feedback is disabled,
indicating that they arise through the coupling between source
redistribution, the ionisation field, and subsequent galaxy evolution.
The recalibrated morphology is also only weakly affected. The fiducial MFP
distributions remain close to the baseline. In the high-mass-dominated
family, \(f_{\rm esc}\) scatter shifts probability towards shorter paths,
consistent with saturation at the upper bound \(f_{\rm esc}=1\). In the
low-mass-dominated family, removing SFR scatter produces longer paths with
UVB feedback but a weak shift towards shorter paths without it. The
morphological response therefore depends on both the galaxies supplying the
ionising emissivity and their sensitivity to photoheating.

The corresponding 21-cm power is much more sensitive to changes in the
global history than to this residual morphology. Defining \(D\) as the difference
between the 21-cm power of a given model and its baseline, normalised by the
forecast uncertainty, we can quantify the impact of stochasticity on the signal.
At \(k\simeq1\,{\rm cMpc}^{-1}\), the uncalibrated models reach maximum
forecast-normalised differences of \(3.4\lesssim|D|\lesssim26.1\) from
their family baselines over \(5\leq z\leq8\). After recalibration, the
largest difference is \(|D|=1.70\), and every model remains within the
forecast \(2\sigma\) uncertainty of a \(1080\)-h {\it SKA}-Low AA4
observation.

The X-ray calculations show the same hierarchy. Uncalibrated
\(0.5\)-dex X-ray luminosity scatter increases the mean X-ray emissivity,
shifting the global absorption minimum from \(-115\,{\rm mK}\) at
\(z\sim17.5\) to \(-95\,{\rm mK}\) at \(z\sim18.5\), advancing the
absorption--emission transition from \(z\sim11.5\) to \(z\sim13\), and
moving the power minimum from \(z\simeq12\) to \(z\simeq13.5\).
Removing SFR scatter produces the complementary response:
it lowers the mean X-ray and stellar Ly\(\alpha\) emissivities, delaying
Ly\(\alpha\) coupling and shifting the absorption--emission transition and
power minimum to \(z\sim11\). Restoring the mean X-ray emissivity in the
luminosity-scatter model, and both mean emissivities in the no-SFR-scatter
model, recovers the timing of coupling and heating. The global signal then
returns to within \(0.1\,{\rm mK}\) of the baseline. At
\(k\simeq1\,{\rm cMpc}^{-1}\), the recalibrated
power differences remain below \(2\) per cent for the
\(L_{\rm X}\)-scatter model and \(6\) per cent for the
no-SFR-scatter model over \(10\lesssim z\lesssim20\). Both
lie within the forecast \(1\sigma\) uncertainty of the
\(1080\)-h {\it SKA}-Low AA4 observation throughout this interval.

Taken together, our results show that the main consequence of introducing
or removing scatter is to shift the population-averaged source
emissivities. Once the relevant means are restored, the timing of
reionization, Ly\(\alpha\) coupling, and IGM heating closely follows the
baseline, while the residual differences in ionisation morphology and
21-cm power are small and model dependent. These residuals remain sensitive
to which halo masses dominate the emissivity and to their coupling with
feedback, but are subdominant for the prescriptions and scales considered
here. For these observables, calibrating the mean source emissivities is
therefore more important than resolving the adopted galaxy-to-galaxy
scatter.
\section*{Acknowledgements}

The simulations presented in this work were carried out using Gadi at the National Computational Infrastructure (NCI Australia) and the OzSTAR supercomputing facility at Swinburne University of Technology. We acknowledge the Australian Government's support for NCI through the National Collaborative Research Infrastructure Strategy (NCRIS). The OzSTAR program is supported by the Australian Government through the Astronomy NCRIS allocation and by the Victorian Government through the Victorian Higher Education State Investment Fund (VHESIF). YQ is supported by the ARC Discovery Early Career Researcher Award (DECRA) through fellowship \#DE240101129. The authors thank S.Wyithe for his constructive comments on draft versions of this work. OpenAI's ChatGPT and Codex were used to assist with language revision and the development and verification of the analysis workflow. All scientific decisions, validation of the results, and responsibility for the final manuscript remain with the authors.

\section*{Data availability}

The data underlying this article will be shared on reasonable request to the corresponding author.
\bibliographystyle{mnras}
\bibliography{references}

\appendix

\section{Uncalibrated ionisation morphology}
\label{app:uncalibrated_morphology}

\begin{figure*}
    \centering
    \includegraphics[width=0.9\linewidth,height=0.82\textheight,keepaspectratio]
    {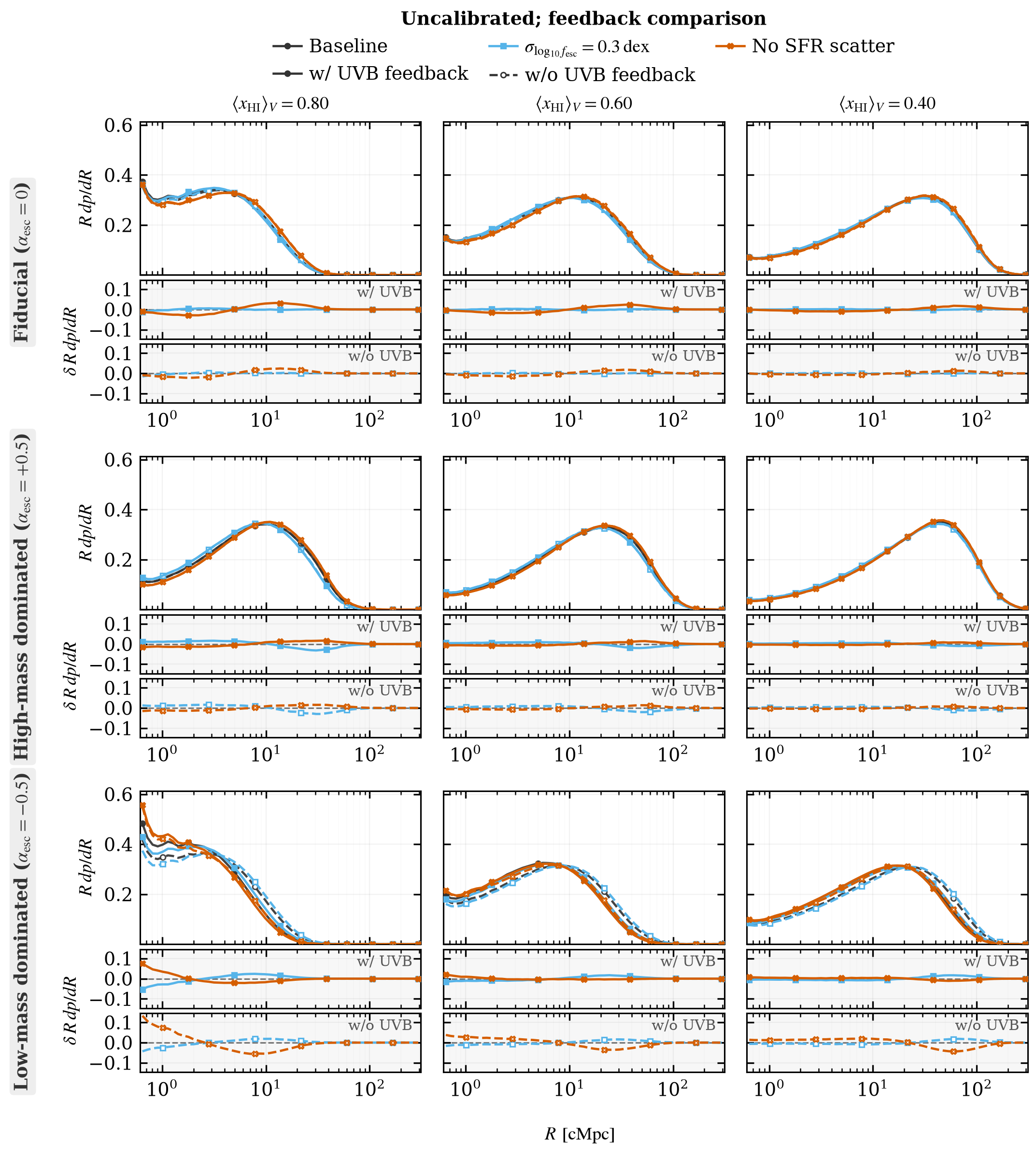}
    \caption{Similar to Fig.~\ref{fig:mfp_recalibrated}, but for the uncalibrated source modifications.}
    \label{fig:mfp_uncalibrated}
\end{figure*}

The main analysis uses the recalibrated models to isolate morphological
changes caused by redistributing the escaped ionising emissivity. For
completeness, Fig.~\ref{fig:mfp_uncalibrated} shows the corresponding MFP
distributions before recalibration. Because the source modifications also
change the mean emissivity, models compared at fixed
\(\langle x_{\rm HI}\rangle_{\rm V}\) can reach that stage at different
redshifts. Their differences therefore include the combined effects of
source redistribution and the shifted reionization history.

\bsp
\label{lastpage}
\end{document}